\documentclass{emulateapj}
\usepackage{amsmath, amsthm, amssymb} 
\usepackage{natbib}
\shorttitle{Neptune on tiptoes}
\shortauthors{Wolff et al.}

\newcommand{\gkbo}{g_{\rm KBO}}
\newcommand{\efree}{e_{\rm free}}
\newcommand{\ifree}{i_{\rm free}}
\newcommand{\eforced}{e_{\rm forced}}
\newcommand{\iforced}{i_{\rm forced}}
\newcommand{\alpharat}{\frac{b_{3/2}^{(2)}(\alpha)}{  b_{3/2}^{(1)}(\alpha) }}

\begin{document}

\title{Neptune on tiptoes:\\ dynamical histories that preserve the cold classical Kuiper belt}
\slugcomment{In press ApJ; accepted: November 29, 2011}

\author{Schuyler Wolff} 
\affil{Johns Hopkins University \\
   3400 N. Charles Street, Baltimore, MD 21218 USA}
\affil{Western Kentucky University \\
   1905 College Heights Blvd., Bowling Green, KY 42101 USA}
   \email{swolff@pha.jhu.edu}
\author{Rebekah I. Dawson}
          \email{rdawson@cfa.harvard.edu}
\author{Ruth A. Murray-Clay}
          \email{rmurray-clay@cfa.harvard.edu}
\affil{Harvard-Smithsonian Center for Astrophysics \\ 
       60 Garden Street, Cambridge, MA 02138 USA}

\begin{abstract}
The current dynamical structure of the Kuiper belt was shaped by the orbital evolution of the giant planets, especially Neptune, during the era following planet formation, when the giant planets may have undergone planet-planet scattering and/or planetesimal-driven migration. Numerical simulations of this process, while reproducing many properties of the belt, fail to generate the high inclinations and eccentricities observed for some objects while maintaining the observed dynamically ``cold" population. We present the first of a three-part parameter study of how different dynamical histories of Neptune sculpt the planetesimal disk. Here we identify which dynamical histories allow an \emph{in situ} planetesimal disk to remain dynamically cold, becoming today's cold Kuiper belt population. We find that if Neptune undergoes a period of elevated eccentricity and/or inclination, it secularly excites the eccentricities and inclinations of the planetesimal disk. We demonstrate that there are several well-defined regimes for this secular excitation, depending on the relative timescales of Neptune's migration, the damping of Neptune's orbital inclination and/or eccentricity, and the secular evolution of the planetesimals. We model this secular excitation analytically in each regime, allowing for a thorough exploration of parameter space. Neptune's eccentricity and inclination can remain high for a limited amount of time without disrupting the cold classical belt. In the regime of slow damping and slow migration, if Neptune is located (for example) at 20 AU, then its eccentricity must stay below 0.18 and its inclination below 6$^\circ$.
\end{abstract}


\section{Introduction}
\label{sec:intro}

The solar system is often used as a case study for the formation of planetary systems from proto-planetary disks. The current configuration of Kuiper belt objects (KBOs) provides a map for how the dynamical evolution of the giant planets in our solar system sculpted the disk of planetesimals.  Therefore it is possible to use the orbital properties of the Kuiper belt to constrain how the orbits of the giant planets evolved in the early solar system, particularly for Neptune, the primary sculptor of the Kuiper belt. Models employing N-body integrations to trace the effects of the giant planets' migration and orbital eccentricity evolution on the planetesimal disk have enjoyed substantial success in reproducing the dynamical populations of KBOs observed today \citep[e.g.][]{1993M,1995M,1999H,2003G,2005H,2008L,2008M}. These populations include objects near orbital resonance with Neptune, objects scattering off Neptune, and ``classical" objects decoupled from Neptune. \citep[See][for definitions of the dynamical classes.]{2008G} Yet substantial discrepancies still exist between the simulations and observations, particularly for the classical population. The bias-corrected inclination distribution of observed classical KBOs is bimodal \citep{2001B,2010G,2011V}. The low-inclination, dynamically ``cold" component and the high-inclination, dynamically ``hot" component also have distinct physical properties, including colors \citep{2000T,2003T,2008P}, sizes \citep{2001L,2010F}, albedos \citep{2009B}, and binary fractions \citep{2006S,2008N}. To date no simulations have been able to produce both the high and low inclination classical objects while qualitatively matching their observed eccentricity distribution.

Several theories of the dynamical history of the giant planets in our solar system have been proposed, inspired by the dynamical populations within the Kuiper belt.  One of the most widely accepted models, the Nice Model, stems from a postulated large scale instability in the early solar system \citep[e.g.][]{1999T,2003G,2008M}. Inspired by the Nice Model, \citet{2008L} proposes a scenario in which Neptune is scattered outward onto an eccentric orbit to near its current location from a formation location closer to the sun \citep[e.g.][]{1999T,2002T}. In this scenario, Neptune's eccentricity subsequently damps due to dynamical friction with a remnant disk of planetesimals. This model reproduces the observed resonant population, scattered population, and the hot classical population, but, we argue, does not satisfactorily match the low eccentricities of the observed population of dynamically cold objects in the classical region. Furthermore, it does not produce a sufficient number of high-inclination objects. Other incarnations of the Nice model \citep[e.g.][]{2003G,2008M} include an \emph{in situ} population of cold objects, but, over the course of Neptune's evolution, these objects become excited to higher eccentricities. Whether or not the particular history described by the Nice Model is correct, the prevalence of large eccentricities among extrasolar giant planets suggests that large-scale orbital excitation is common during the formation of planetary systems.

In contrast to the upheaval of the Nice Model, another model \citep{1993M} proposes a period of extensive, smooth migration of the giant planets. Planetesimal-driven migration \citep{1984F} is likely important in shaping the architecture of many planetary systems.  \citet{1993M,1995M,1999H} demonstrated that the migration of the giant planets in our solar system on flat, circular orbits would have perturbed the disk of planetesimals, scattering some to more eccentric orbits and capturing others into resonance. They also found that migration results in a large number of planetesimals in the location known today as the ``Scattered Disk," which encompasses those objects that have had gravitational interactions with one or more of the giant planets and have high orbital eccentricities and inclinations. The key inconsistency between this model and current observations is that it cannot produce, without additional processes such as stochasticity \citep[e.g][]{2003L,2006M}, both the cold and hot classical populations from a single set of initial disk conditions or account for the differences in their physical properties.

Because detailed N-body simulations are computationally expensive, previous works have investigated a limited number of solar system planetary histories. Currently, no comprehensive parameter study has been done exploring the evolution of Neptune's orbit given constraints from the sculpting of the Kuiper belt. In addition, although planet-planet scattering often produces large mutual inclinations \citep{2011C}, as in observed in the Upsilon Andromedae system \citep{2010M}, no serious, detailed treatment has included the possibility that Neptune underwent a period of high orbital inclination, a conceivable outcome of the orbital instability in the early solar system. Here we consider a general model that can encompass the two detailed models described above -- the Nice-Model-inspired scenario in which Neptune is scattered to a high eccentricity \citep{2008L} and the scenario of extensive migration of Neptune on a low-eccentricity, low-inclination orbit \citep[e.g.][]{1995M}  -- as well as potential scenarios in which Neptune undergoes a period of high inclination. In this generalized model, Neptune undergoes some combination of migration and/or evolution of its eccentricity and/or inclination. Here we take a step toward understanding the qualitative differences in the dynamics of the Kuiper belt generated by a wide range of planetary histories. In a series of three papers, we perform a parameter study of the effects on a disk of planetesimals of the migration and the eccentricity and inclination evolution of Neptune in the early solar system. No component of this parameterization is new; rather, our goal is to comprehensively consider all possible parameters for Neptune's history.

In this first paper, we introduce our method for computationally and analytically modeling Neptune's orbital evolution and its effects on the planetesimal population. Then we demonstrate several concepts that will allow us to thoroughly explore the parameter space of Neptune's dynamical history: 
\begin{itemize}
\item If Neptune undergoes a period of elevated orbital eccentricity, it will secularly excite the eccentricities and inclinations of an \emph{in situ} planetesimal population. We can analytically model this secular excitation using a simple expression.
\item As Neptune's eccentricity and inclination damp, the planetesimals evolve to final eccentricities and inclinations that depend on Neptune's initial eccentricity and inclination and damping timescales.
\item The effects of Neptune's eccentricity and inclination evolution on the planetesimals can be treated separately to first order.
\item The migration rate sets Neptune's effective location for the secular evolution of the planetesimals.
\end{itemize}
Having demonstrated these points, we can place robust constraints on Neptune's semi-major axis, eccentricity, and inclination during its late evolution (after any period of planet-planet scattering), and on its migration, eccentricity damping, and inclination damping rates, identifying which parameters are consistent with maintaining the low eccentricities and inclinations of the cold classicals. In the second paper \citep{2012D}, we place even stronger constraints by incorporating additional effects, including the effects of other planets on Neptune's orbital evolution and the effects of proximity to mean-motion resonance with Neptune on the secular excitation of the planetesimals, as well as additional constraints from the hot classical population. A third paper (Dawson and Murray-Clay 2012b, in prep) will focus on the inclinations of the classicals.

In Section \ref{sec:nep}, we present our parameterization of Neptune's orbital evolution. In Section \ref{sec:small}, we discuss the observational constraints that today's cold classical KBOs place on past sculpting of the planetesimal disk and describe our computational and analytical models of the excitation of the planetesimal disk by an inclined and eccentric Neptune. We present the results of the parameter study in Section \ref{sec:results}. In Section \ref{sec:conclusions}, we present our conclusions and describe how the companion papers will expand upon this work.

\section{Modeling Neptune's orbital evolution}
\label{sec:nep}

As a first step toward a comprehensive study of the impact of a wide range of dynamical histories of the outer solar system on the Kuiper belt, we thoroughly explore the parameter space of a general model for Neptune's dynamical history. This model encompasses specific, previously-proposed solar system history models \citep[e.g.][]{1995M,2008L}.  In our parameterization, Neptune is instantaneously scattered to an initial location with an initial eccentricity and inclination. Subsequently, it undergoes planetesimal-driven migration and its eccentricity and inclination are damped by dynamical friction from the disk of planetesimals. Resonant relaxation may also contribute significantly to the planet's eccentricity and inclination damping if the Toomre parameter of the planetesimal disk is large or if the random velocities of the planetesimals are large \citep{1998T}. The parameters of this model are Neptune's ``initial" (defined below) semi-major axis, eccentricity and inclination, the planet's migration rate, and the timescales for Neptune's eccentricity and inclination damping. In this model, we include the effects of only one planet (Neptune), an approach we justify briefly in Section \ref{subsec:justnep} and more thoroughly in \citet{2012D}.

\subsection{Parameters}
\label{subsec:param}
We define Neptune's orbital evolution using the following parameters:
\begin{itemize}
\item An ``initial" semi-major axis, eccentricity, and inclination. The initial semi-major axis is not necessarily the location where Neptune formed. We imagined that Neptune underwent a scattering, or series of scatterings, onto an inclined and eccentric orbit, which happened quickly enough to not disrupt the cold classical belt. We model the period after the scatterings end and Neptune's eccentricity and inclination begin to damp.
\item A migration rate, defined in Section \ref{subsec:modelnep}.
\item An eccentricity damping rate and inclination damping rate, also defined in Section \ref{subsec:modelnep}.
\end{itemize}

\subsubsection{Values for the migration parameters}

Here we consider what range of parameters we should explore for Neptune's migration direction, distance, and timescale:

\begin{itemize}
\item Migration direction: We know that Neptune's migration will be, on average, outward because Neptune is exterior to a very massive planet (in this case Jupiter). As Neptune scatters planetesimals, Jupiter ejects them, leading to a net loss in angular momentum for Jupiter and gain in angular momentum for Neptune \citep[e.g.][]{1984F,1993M,1995M}.
\item Migration distance: If all the KBOs in orbital resonance with Neptune were captured during migration and from orbits with low eccentricity, Neptune needs to have migrated a distance of 7-10 AU \citep{1993M,1995M,2005H} to adiabatically raise their eccentricities to the values observed. However, other mechanisms have been proposed for producing the population of resonant KBOs, in which case the 7-10 AU constraint on the migration distance would not apply. \citet{2008L} argue that resonant objects were scattered from the inner disk into the classical region, entered resonances widened by Neptune's high eccentricity, and were trapped when Neptune's eccentricity damped. In a companion paper \citep{2012D}, we demonstrate that objects scattered into classical region secularly evolve very quickly near resonances, allowing them to reach stable, lower-eccentricity orbits. Future observations of the binary fractions (and possibly colors) of resonant KBOs -- properties that \cite{2011MS} argue may encode the location of each KBO's formation -- may distinguish between these mechanisms. Therefore we consider all ``initial" semi-major axes for Neptune interior to its current location. We quote constraints for 20 AU and 30 AU as examples of a long and short migration distance, respectively.
\item Migration timescale: The distribution of libration angles of twotinos, KBOs in the 2:1 resonance, place a lower limit of 1 Myr on Neptune's migration timescale \citep{2005M}. (Upcoming unbiased surveys, e.g. PAN-STARRS, LSST, are needed to confirm the true distribution of the twotinos.) An upper limit could be imposed by the stochasticity of the migration, which due to the finite sizes of planetesimals, limits the efficiency of keeping objects in resonance \citep{2006M}. However, the amount of stochasticity depends on the size distribution of the planetesimals, which is unknown and depends on the physics of their formation \citep[see][and references therein]{2010C}. Given these uncertainties, we consider all migration timescales greater than 0.3 Myr.
\end{itemize}

\subsection{Computational model}
\label{subsec:modelnep}
Direct computational modeling of the effect of planetesimals on Neptune's orbit would be computationally expensive, so instead we apply fictitious forces (Appendix) to evolve Neptune's semi-major axis $a_N$, eccentricity $e_N$, and inclination $i_N$, with any specified functional form. Following \citet{1993M} and \citet{2008L}, we use the functional forms:

\begin{eqnarray}
\label{eqn:forms}
e_N = e_0 \exp{(-t/\tau_e)} \nonumber\\
i_N = i_0 \exp{(-t/\tau_i)} \nonumber\\
a_N = a_f + (a_0 - a_f) \exp{(-t/\tau_a)}
\end{eqnarray}
\noindent where $a_0$ is the initial semi-major axis of Neptune, $a_f$ = 30 AU is the final semi-major axis, and $\tau_e$, $\tau_i$, and $\tau_a$ are the eccentricity damping timescale, inclination damping timescale, and migration timescale respectively. Our results do not depend on the specific form of Eqn (\ref{eqn:forms}). As we will demonstrate in Section \ref{sec:results}, sometimes the instantaneous rate of change of the variables ($\frac{\dot{a}}{a}, \frac{\dot{e}}{e}, \frac{\dot{i}}{i}$) is most relevant, while in other cases the total evolution matters most. We have verified these statements with integrations (not shown) using an alternative migration form $\frac{\dot{a}}{a} \varpropto \frac{\dot{e}}{e} \varpropto \frac{\dot{i}}{i} \equiv {\rm constant}$.

\section{Planetesimals: observational constraints and modeled evolution}
\label{sec:small}

We model an initially unexcited disk of planetesimals that becomes today's cold classical population. In Section \ref{sec:obs}, we present the observational constraints on the excitation of this population. In Section \ref{subsec:sec}, we present an analytical model for the evolution of this planetesimal disk under the influence of Neptune, which we use to predict and interpret the results of numerical simulations. In Section \ref{subsec:justnep}, we justify directly modeling only Neptune instead of all four giant planets.

\subsection{Constraints from the observations of cold classical objects}
\label{sec:obs}

The cold classicals are a class of dynamically ``cold" objects on low-eccentricity, low-inclination orbits, with positions starting at 42.5 AU, the region interior to which is unstable, and falling off quickly beyond 45 AU \citep{2009K}. We assume that today's cold classical KBOs are remnant planetesimals that formed \emph{in situ}, and we use these terms interchangeably.  Strong constraints can be placed on the dynamical history of the solar system by requiring that Neptune not disrupt these objects as it migrates outward on an inclined and/or eccentric orbit. In this section, we discuss our specific criteria for ``preserving" the cold classicals, which we will use to place constraints on the set of parameters (Section \ref{subsec:param}) defining Neptune's orbital history.

\begin{figure}[htbp]
\begin{centering}
\includegraphics[width=.45\textwidth]{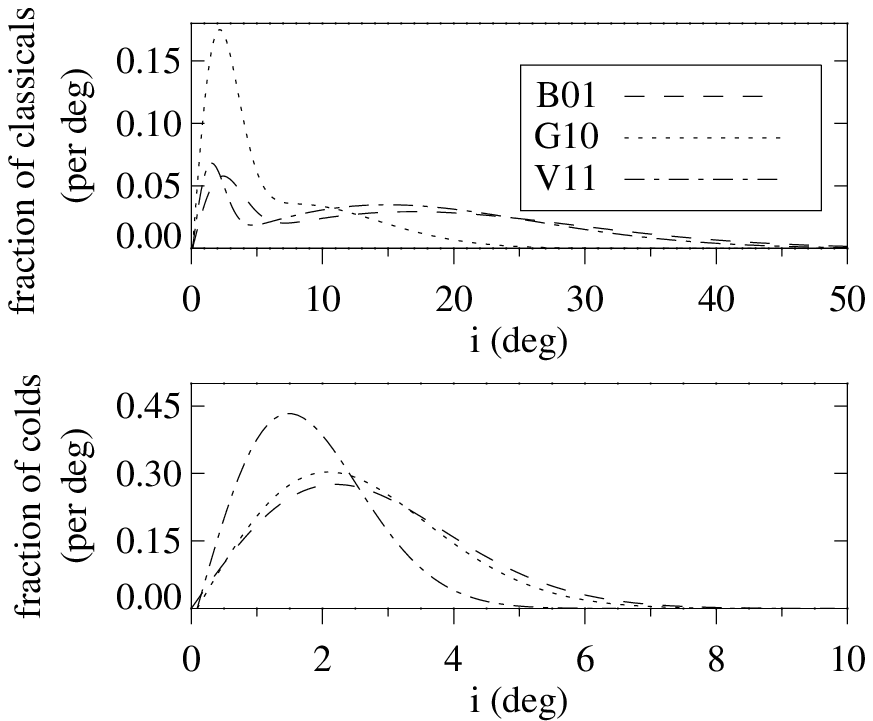}
\caption{Three models of the un-biased inclination distribution of the classicals: \citet{2001B}, dashed line; \citet{2010G}, dotted line; and \citet{2011V}, dot-dashed line.  The biggest difference among them is that in the \citet{2010G} model, the hot (high inclination) component is a much less substantial portion of the total classical population and has a substantially smaller inclination width than in the other two models. The distributions are very similar for the inclinations of the cold classicals (bottom panel). \label{fig:distcompare}}
\end{centering}
\end{figure}

There is evidence that the classical KBOs have a bimodal inclination distribution \citep{2001B,2010G,2011V}. The cold classicals are defined as the class of objects with a distribution of inclinations $i$ centered on a low inclination with a small width in inclination. The functional form of this distribution is typically modeled as a Gaussian multiplied by $\sin i$. One \citep{2010G} of the three proposed models for the de-biased inclination distribution differs substantially from the other two \citep{2001B,2011V} with respect to the relative populations of the cold and hot classicals and the width of the hot (high $i$) component (Fig. \ref{fig:distcompare}, top panel). However, all three distributions are similar for the cold classicals (Fig. \ref{fig:distcompare}, bottom panel). The number of cold classicals per inclination bin falls off almost entirely by $i = 6^\circ$ for the models of \citet{2001B} and \citet{2010G} and by $i = 4^\circ$ for the model of \citet{2011V}. Therefore we require that Neptune's dynamical history should not excite the cold classicals above an inclination of 6 degrees.

The disruption criterion for the cold classical KBO eccentricities is more subtle. The cold population is defined by its inclination distribution, not its eccentricity distribution. We could imagine a cold population of objects which have inclinations below six degrees but a uniform distribution of eccentricities. If this were true (and we will demonstrate that it is not), the initial planetesimal disk could be excited to arbitrarily large eccentricities. Moreover, the eccentricity distribution could be shaped entirely by the long-term stability of the KBO orbits. In this case, objects excited in eccentricity during Neptune's high-eccentricity period would be ejected from the system over billions of years. 

\begin{figure*}[htbp]
\begin{center}
\includegraphics[width=\textwidth]{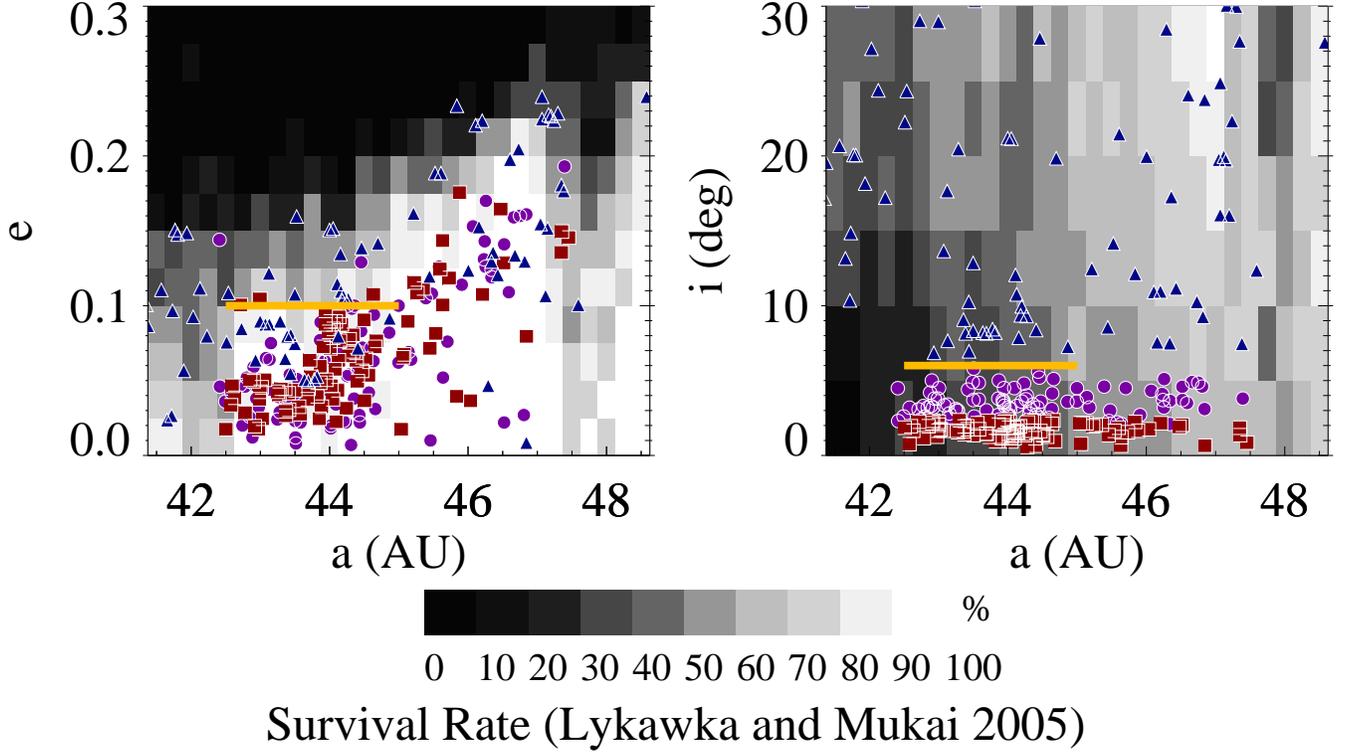}
\caption{Plotted over the survival maps of \citet{2005L} are the eccentricity (left) and inclination (right) distributions of the observed classical KBOs. The red squares are objects with $i < 2^\circ$ and are thus very likely cold classicals. The blue triangles have $i > 6^\circ$ degrees and are thus very likely hot classicals. The membership of any given purple circle ($2^\circ < i <  6^\circ$) is ambiguous. In eccentricity, the cold classicals (red squares) between 42.5-44 AU are confined to $e < 0.05$, well below the survival limit, while cold classicals between 44-45 AU are confined to $e < 0.1$, also below the survival limit in this region. Classical objects are taken from the Minor Planet Center Database and classified by \citet{2008G} and \citet{2011V}. The yellow lines indicate the conservative criteria for preserving the cold classicals. \label{fig:diagnostic}}
\end{center}
\end{figure*}

To test whether the eccentricities of the cold classicals are sculpted solely by stability, we compared observed cold classical objects to a stability map created by \citet{2005L} (Fig. \ref{fig:diagnostic}). \citet{2005L} do not use proper elements, so we use the instantaneous orbital elements of the observed objects. We have confirmed that the features of the distributions we identify below are qualitatively the same using proper elements \citep[][Appendix A.1]{2012D}. Because the hot and cold population overlap (Fig. \ref{fig:distcompare}), we cannot definitively determine to which distribution any particular object belongs. However, for all three model distributions, less than 10$\%$ of objects with inclinations $i<2^\circ$ are hot. We find that the eccentricities of $i<2^\circ$ objects in the region from 42.5-45 AU are confined well below the survival limit. Consequently, we can conservatively constrain the dynamical history of Neptune: Neptune cannot excite the cold classical objects in this region above e = 0.1. 

Thus we can impose two conservative criteria for preserving the cold classicals: 
\begin{enumerate}
\item In the region from 42.5 to 47.5 AU, the inclinations of the cold classicals must not be excited above $i < 6^\circ$
\item In the region from 42.5 to 45 AU, the eccentricities must not be excited above $e < 0.1$.
\end{enumerate}

\subsection{Secular evolution model for the planetesimals}
\label{subsec:sec}

Excitation of the \emph{in situ} planetesimal population by a perturbing planet on an inclined and/or eccentric orbit occurs through secular evolution. Here we present simple analytical expressions that we will use to predict and interpret the results of our integrations in Section \ref{sec:results}.

\subsubsection{Dynamics of secular evolution}

Due to forcing from Neptune\footnote{Other planets besides Neptune contribute to the secular forcing, but as a simplification we consider only the effects of Neptune. See Section \ref{subsec:justnep} for a detailed justification.}, both the eccentricity and inclination of a planetesimal undergo secular oscillations on timescales of order a million years. The planetesimal's total eccentricity is the vector sum of its forced eccentricity $\eforced$, imparted by Neptune, and its free eccentricity $\efree$, set by initial conditions (Fig. \ref{fig:forcefree}). The $\efree$ vector precesses about the $\eforced$ vector at the angular frequency $\gkbo$. The secular evolution of the planetesimal's inclination is analogous to -- yet, to lowest order, separable from -- the eccentricity evolution. The planetesimal's total inclination is the sum of its forced and free inclination, and the free inclination precesses about the forced inclination. In the case of inclination, we can think of the particle's orbit being inclined by $\ifree$ with respect to the ``forced" plane and precessing about the forced plane at the rate $\gkbo$. See \citet{2000M}, Chapter 7, for a pedagogical presentation of secular evolution.

\begin{figure}[htbp]
\begin{center}
\includegraphics[width=.45\textwidth]{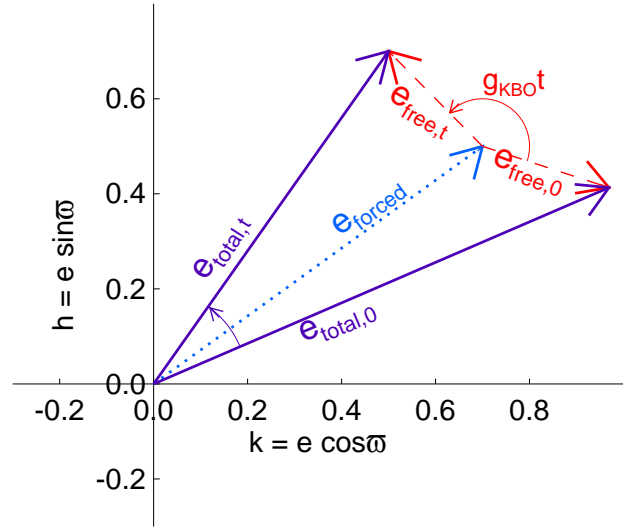}
\caption{At a given time $t$, the total eccentricity (purple, solid) is the vector sum of $\eforced$ (blue, dotted) and $\efree$ (red, dashed). The vector $\efree$ precesses about $\eforced$ at the rate $\gkbo$, so at time $t$, the $\efree$ vector has rotated by an angle $\gkbo t$ from its initial orientation with respect to $\eforced$. \label{fig:forcefree}}
\end{center}
\end{figure}

The vector components of the planetesimal's eccentricity are $h = e \sin \varpi$ and $k = e \cos \varpi$, where $\varpi$ is the planetesimal's longitude of periapse. Secular forcing by Neptune causes $h$ and $k$ to evolve as (to first order in $e$ and $e_N$):

\begin{eqnarray}
\label{eqn:sece}
h = \efree \sin(\gkbo t + \beta) + \eforced  \sin (\varpi_{N}) \nonumber \\
k = \efree \cos(\gkbo t + \beta) + \eforced \cos (\varpi_{N}) \nonumber \\
\end{eqnarray}
where 
\begin{eqnarray}
\label{eqn:simples}
\eforced =   \alpharat e_{N}, \nonumber \\
\alpha = \frac{a_{N}}{a}, \nonumber \\
g _{\rm KBO} = \alpha b_{3/2}^{(1)}(\alpha)  \frac{m_N}{m_{\rm sun}}   \frac{n}{4}   \nonumber \\
\end{eqnarray}

The constants $\efree$ and $\beta$ are determined from the initial conditions. Here, $\varpi_N$ is the longitude of periapse of Neptune, $e_N$ is the eccentricity of Neptune, and $\alpha$ is the ratio of Neptune's semi-major axis to that of the planetesimal, $a$, all of which are assumed to be constant. The functions $b$ are standard Laplace coefficients. The secular frequency of the KBO is $\gkbo$, $m_N$ is the mass of Neptune, $m_{\rm sun}$ is the mass of the Sun, and $n = (Gm_{\rm sun}/a^3)^{1/2}$ is the planetesimal's mean motion. 

Similarly, the vector components of the planetesimal's inclination are $p = i \sin \Omega$ and $q = i \cos \Omega$, where $\Omega$ is the planetesimal's longitude of ascending node. Secular forcing by Neptune causes $p$ and $q$ to evolve as (to first order in $i$ and $i_N$):
\begin{eqnarray}
\label{eqn:seci}
q = \ifree \sin(-\gkbo t + \gamma) + \iforced \sin (\Omega_{N}) \nonumber \\
p = \ifree \cos(-\gkbo t + \gamma) + \iforced \cos (\Omega_{N}) \nonumber \\
\end{eqnarray}
\noindent where
\begin{eqnarray}
\label{eqn:simplesi}
\iforced = i_N
\end{eqnarray}
The constants $\ifree$ and $\gamma$ are determined from the initial conditions. Here, $\Omega_N$ is the longitude of ascending node of Neptune and $i_N$ is the inclination of Neptune. While $\eforced$ depends on $\alpha$ (Eqn. \ref{eqn:simples}), to first order $\iforced$ is equal to $i_N$, regardless of the particle's semi-major axis. 

Three quantities are plotted in Fig. \ref{fig:freqs} as a function of $a$ (for two different $a_N$): 1) $\eforced/e_N$, 2) $\iforced/i_N$, and the secular period $\frac{2\pi}{\gkbo}$. These values describe the amplitude and timescale of the secular excitation of the \emph{in situ} planetesimal population. For example, when Neptune is at 20 AU, a planetesimal at 45 AU has a forced eccentricity of $0.55 e_N$ and a forced inclination of $i_N$. If the planetesimal begins with $e = i = 0$, it will reach its maximum eccentricity $e = 2\times0.55 e_N = 1.1 e_N$ and maximum inclination $i = 2 i_N$ on a timescale of $\frac{1}{2} \frac{2\pi}{\gkbo}$ = 13 Myr (i.e. half a secular oscillation period).

\begin{figure}[htbp]
\begin{center}
\includegraphics[width=.45\textwidth]{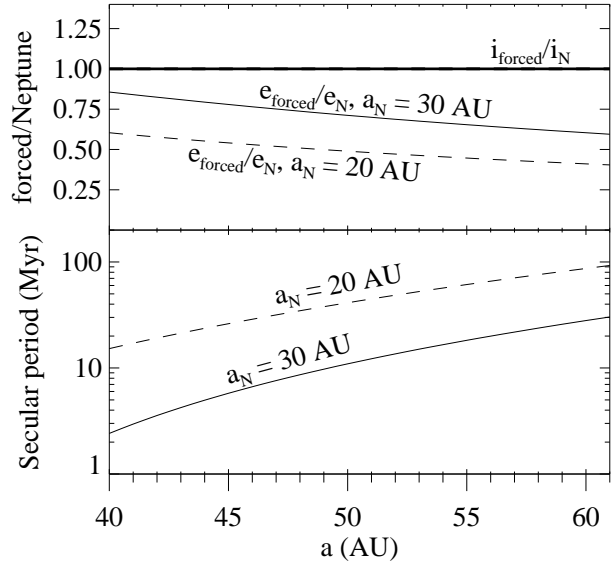}
\caption{Parameters, to first order, governing secular evolution -- using two different locations of Neptune, 20 AU (dashed line) and 30 AU (solid line) -- as a function of the planetesimal's semi-major axis. Top: Ratio of the planetesimal's forced eccentricity (dashed, solid) and forced inclination (thick, same for $a_N = 20$ AU, $a_N = 30$ AU) to that of Neptune. To first order, the forced inclination does not depend on the position of the planetesimal relative to Neptune, while the forced eccentricity decreases with the planetesimal's semi-major axis. Bottom: Timescale of the secular evolution.\label{fig:freqs}}
\end{center}
\end{figure}

\subsubsection{Secular excitation of the planetesimal disk}

The secular excitation of a planetesimal disk can be modeled using the expressions above. Here we consider that Neptune is temporarily on an inclined and/or eccentric orbit after undergoing scattering on an effectively instantaneous timescale (i.e. much less than the secular timescale). Then Neptune imparts a forced eccentricity $\eforced$ and inclination $ \iforced$ on the initially cold planetesimal disk. Thus each planetesimal has $\efree \sim \eforced$. and $\ifree \sim \iforced$. The planetesimal's total eccentricity and inclination, each a vector sum of the free and forced, now oscillates from the initial values of $e(0) \sim 0$ and $i(0) \sim 0$, reaching maximum values of $e = \efree +  \eforced \sim 2 \eforced$ and $i = \ifree +  \iforced \sim 2 \iforced$, on a timescale set by the secular evolution rate $\gkbo$. Thus an initially cold planetesimal disk will become excited.

\subsection{The case for modeling only Neptune's effects on the planetesimals}
\label{subsec:justnep}

We limit our parameter study to include only the planet Neptune. We have several reasons for choosing this approach. First, unlike previous approaches, we are not attempting to create a single model that reproduces the entire Kuiper belt in detail, but to constrain which histories of Neptune are consistent with a major qualitative feature: the unexcited orbits of the cold classicals. Our constraints will feed into more detailed models. Second, restricting the parameter study to just Neptune drastically reduces the number of parameters, allowing us to thoroughly explore the remaining parameter space. Third, we find \citep{2012D} that the primary effects of the other planets on the Kuiper belt are indirect: they alter the orbit of Neptune, which in turn affects the Kuiper belt. Our modifications to the \emph{Mercury 6.2} integrator (Appendix A) allow us to model any orbital evolution of Neptune without needing to include the other planets. Fig. \ref{fig:otherplanets} provides one example pair of integrations, demonstrating that cold planetesimals evolve similarly in the presence of only Neptune and in the presence of all four giant planets. 

\begin{figure}[ht!]
\begin{centering}
\includegraphics[width=.45\textwidth]{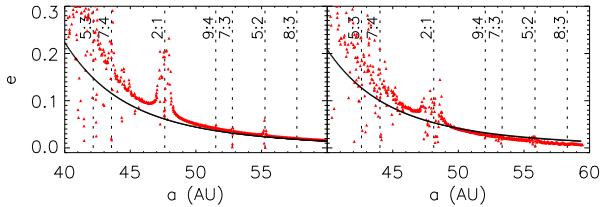}
\caption{{Snapshot at 0.5 Myr of test particles under the influence of Neptune alone (left) and all four giant planets (right). In both cases, the planetesimals begin at $e=0$ and Neptune has $a_N = 30, e_N = 0.2$. Saturn, Jupiter, and Uranus are included in the integration on their current orbits (right) as an illustrative case. The black line is the same in both panels: the first-order predicted secular excitation of the planetesimals under the influence of just Neptune (Eqn. \ref{eqn:sece}), neglecting the effects of resonances. Outside of the resonances, the results are qualitatively similar. In this case (right), the forced precession period of Neptune due to the other giant planets (approximately 4.5 Myr) slow or comparable to the timescale of planetesimal secular evolution. Thus Neptune's precession has a negligible effect on the secular evolution of the planetesimals.}\label{fig:otherplanets}}
\end{centering}
\end{figure}

One main effect of other planets is to cause Neptune's longitude of periapse to precess. However, the disruption of the cold classicals is not significantly affected if Neptune's precession period is comparable to or longer than the secular excitation time (Fig. \ref{fig:otherplanets}). If Neptune precesses very quickly, the cold classicals could be preserved outside our constraints, as proposed by \citet{2011B}. In addition, strong interactions between Neptune and Uranus can cause Neptune's semi-major axis to oscillate, leading to orbital chaos in the classical region, which can cause additional excitation. We explore these additional complications in \citet{2012D}.

Here we do not explore the $g_8$ and $\nu_8$ secular resonances, which over time remove low eccentricity and low inclination objects respectively. 
The location of these secular resonances will impose additional constraints on preserving the cold population. However, since they \emph{remove} unexcited objects, leaving excited objects behind, these resonances will not allow for parameters of Neptune that we rule out.

\subsection{Considerations for a massive planetesimal disk}

Throughout the paper, we treat the planetesimals as massless test particles. We parametrically model the orbital evolution of Neptune caused by the planetesimals -- Neptune's migration and the damping of its eccentricity and inclination -- but do not explicitly consider the effects of planetesimal self-gravity. In this section, we consider some of these effects and how they might impact our conclusions.

One important consideration is whether the transfer of angular momentum between Neptune and the disk is sufficient to excite a massive disk to the extent we assume for massless test particles. Since the total angular momentum of the system is conserved, the angular momentum deficit (AMD) due to the secular excitation of the cold classicals cannot exceed Neptune's initial angular momentum deficit \citep[See][for a description of the concept of AMD]{1997L}. Using the expansion of the AMD by \citet[Eqn. 26a and 26b]{2003H}, the ratio of the total AMD of the cold objects to the AMD of Neptune is:
\begin{eqnarray}
\frac{\rm AMD_{\rm cold}}{\rm AMD_{\rm Neptune}} = \frac{ m_{\rm cold} }{m_N} \sqrt{\frac{a_{\rm cold}}{a_N}} (\frac{e_{\rm cold}^2+i_{\rm cold}^2}{e_N^2+i_N^2})
\end{eqnarray}
Since the factor $\sqrt{\frac{a_{\rm cold}}{a_N}}$ is roughly unity, if the mass the cold classicals exceeds Neptune's mass, the AMD of Neptune would be too small to excite the cold classicals to $e_N$ and $i_N$. Thus if the mass in cold classicals were large enough, the objects could remain at low eccentricities and inclinations even if Neptune's eccentricity and inclination were large.

The outer solar system likely contained several tens of Earth masses at early times, comparable to Neptune's mass of 17$ M_\oplus$. Such a massive disk is required to form large KBOs such as Pluto by coagulation \citep{1997S,1998K} and to drive substantial migration \citep{1984F,1999H} and/or eccentricity and inclination damping of Neptune's orbit. In contrast, \citet{2008F} combined the results of several surveys to estimate that the mass of the current classical Kuiper belt is only $0.008 \pm 0.001 M_\oplus$, 40$\%$ of which is the cold component \citep{2009K}. 

We assume throughout the rest of this paper that the total mass -- and thus AMD -- of the cold classicals is smaller than that of Neptune (i.e. closer to today's mass). This choice implicitly assumes that the primordial disk was partially truncated or that the cold classicals were depleted before the era we treat in this paper. If the cold classical belt began with a low-mass -- compared to the inner disk, responsible for Neptune's migration and for dynamical friction -- truncation of the planetesimal disk may explain why Neptune's migration halted at 30 AU \citep{2003L}. Dynamical depletion would tend to excite the cold classical population to higher eccentricities than observed, but could be consistent if all excited objects were subsequently scattered out of the region from 42.5-45 AU. Thus this dynamical depletion would need to occur before the delivery of the observed hot classical population, which has higher eccentricities. During the era we consider in this paper, Neptune must deliver the hot objects without disrupting the cold ones.

The constraints presented here will require revision if Neptune's primary sculpting of the Kuiper belt occurred while the disk was massive. For example, this could be true if the cold classical belt was depleted through collisions following Neptune's orbital evolution. Collisional depletion occurs through collisional grinding, followed by ejection of small grains by radiation forces, and \citet{1998K} and \citet{2001K} argue for this scenario in the context of coagulation models for the growth at KBOs. A break in the size distribution of the cold classicals at $R = 25- 50$ km has been attributed to collisional grinding \citet{2005P}. However, \citet{2011N} argue that, given this interpretation, a large fraction of binaries with small components ($R<50$ km) would have been disrupted. Such binaries are observed, implying that collisional grinding did not generate the break. Hence collisional grinding could only have substantially depleted the cold Kuiper belt if the majority of the belt's mass was initially sequestered in much smaller objects.

Another potentially important effect is the propagation of spiral density waves and torsion waves in a massive disk \citep{1998W,2003H}, excited at secular resonances and at the edge of the disk. In particular, spiral density waves tend to ``smear out" the excitation of the eccentricities and inclinations of the planetesimals near secular resonances. Spiral density waves are potentially important in sculpting the dynamical structure of the Kuiper belt --  \citet{1998W} use them to place constraints on the mass of the Kuiper belt beyond 50 AU. The exact behavior of the density waves depends on a variety of disk parameters -- including the size distribution of planetesimals, the surface density profile in the disk, and the thickness of the disk -- so we refer the reader to \citet{2003H} for a detailed exploration. (For ease of comparison, note that assuming a surface density profile $\Sigma \propto a^{-1.5}$, the current mass of the cold Kuiper belt corresponds to a $\Sigma$ roughly a factor of 3 smaller than the lowest surface density simulated by \citet{2003H}.) Without the influence of density waves, very large forced eccentricities and inclinations ($e \rightarrow \infty, i \rightarrow \infty$) occur only close to secular resonances. The density waves spread the elevated eccentricity and inclination, confined to the secular resonance in the massless disk case, across the Kuiper Belt from 40-50 AU. However, our constraints do not include the extra excitation caused by secular resonances, which excite the cold classicals even more than we consider. Therefore, the constraints we will place on \emph{ruling out} parameters of Neptune that cause excessive excitation will still hold and, as future work, the additional excitation caused by secular resonances plus density waves may place additional constraints. 

In the absence of secular resonances, waves launched at the disk edge could similarly smear eccentricities and inclinations.  We assume that the eccentricities and inclinations of planetesimals are not self-damped by spiral density waves launched at the disk edge. The eccentricities and inclinations of planetesimals could also be damped by dynamical friction caused by smaller, unobserved bodies whose random velocities were damped by collisions \citep{2004G} or by resonant relaxation \citep{1998T}. We assume that neither of these effects contribute significantly during the era of interest.

Finally, the eccentricities and inclinations of the planetesimals may not be excited in the first place if the planetesimal disk is massive enough to cause Neptune's eccentricity to quickly precess, lowering the forced eccentricity of the cold classicals. In order for the disk to significantly affect the precession rate of the cold classicals, its mass must be at least that of Neptune.  A scenario involving a quickly precessing Neptune is explored in \citet{2011B} and discussed in our companion paper, \citet{2012D}. 

\section{Results}
\label{sec:results}

We present here the results of a set of integrations containing Neptune and a collection of test particles designed to represent the \emph{in situ} planetesimals that become today's cold classical KBOs. The orbital evolution of Neptune is modeled as described in Section \ref{subsec:modelnep} and the Appendix. In Section \ref{subsec:compmodel}, we describe how we computationally model the planetesimals. In Section \ref{subsec:evolve}, we characterize the secular excitation of the planetesimals under the influence of an inclined and eccentric Neptune and place limits on the eccentricity and inclination of Neptune. In Section \ref{subsec:damp}, we determine the effect of damping of Neptune's eccentricity and inclination on the excitation of the planetesimals. We demonstrate that there are two regimes for damping Neptune's orbit -- in the slow-damping regime, the planetesimals evolve to their initial free eccentricity and inclination, set by Neptune's initial eccentricity and inclination, and in the fast-damping regime, their eccentricity and inclination are frozen at the value reached at the damping time. We also show that the evolution of the inclination and eccentricity can be treated separately.  In Section \ref{subsec:migrate}, we consider the effects of Neptune's migration. Finally, we present two example scenarios in Section \ref{subsec:ex} demonstrating the principles we have derived: a pathological scenario in which the evolution of the Neptune's semi-major axis, eccentricity, and inclination occur each on a different timescale, and a realistic scenario consistent with preserving the cold population.

\subsection{Computational model of planetesimals}
\label{subsec:compmodel}

We performed N-body integrations of an initially cold planetesimal disk under the influence of a dynamically-evolving Neptune using the \emph{Mercury 6.2} hybrid symplectic integrator \citep{1999C} with an accuracy parameter of  $10^{-12}$, a step size of 200 days, and user-defined forces and velocities imposing the migration and damping of Neptune, a complete description of which is provided in Section \ref{subsec:modelnep} and the Appendix. We modeled the initial population of planetesimals, which become today's cold classical KBOs, as 600 massless test particles with initial $a$ evenly spaced between 40 to 60 AU and initial $e = i = 0$. We will use the terms ``planetesimals" and ``test particles" interchangeably from here forward. Although simulations that deliver KBOs into the classical region via scattering require $\sim 10^{4}$ planetesimals in each run, our immediate goal of understanding the conditions under which the cold classical population can be retained can be achieved with a smaller number, allowing us to explore a wider range of orbital conditions. 

Integrations probing solar system histories are typically run for 4 Gyr in order to test the stability of the system under current solar system conditions. However, the dynamical histories of Neptune we test are not dependent on the long-term survival of the planetesimals, because we have chosen observational criteria that are independent of the long-term evolution of the belt under the current configuration of the solar system (Secton \ref{sec:obs}). To save computation time, our integrations probe only the period in which Neptune undergoes planetesimal-driven migration and/or eccentricity damping and/or inclination damping.

\subsection{Secular excitation of planetesimals under the influence of an inclined and eccentric Neptune}
\label{subsec:evolve}

In scenarios of the early solar system in which Neptune is scattered onto an eccentric and/or inclined orbit, Neptune can potentially disrupt an \emph{in situ} population of planetesimals above the confined eccentricities and inclinations we observe (Section \ref{sec:obs}) in the cold classical Kuiper belt today. Neptune forces the planetesimals to undergo secular evolution (Section \ref{subsec:sec}), in which they oscillate through high eccentricities and inclinations. The rate of secular evolution depends on the location of the planetesimal relative to Neptune (Fig. \ref{fig:freqs}). Fig \ref{fig:excite} shows snapshots of the secular evolution of test particles under the influence of Neptune when the planet is stationary at 20 AU (top), migrates from 20 to 30 AU (middle), and is stationary at 30 AU (bottom). In their secular evolution, the planetesimals reach maxima $e = 2 \eforced$ and $i = 2 \iforced$. In the middle panel, the evolution transitions from matching the top panel (early snapshots, left) to matching the bottom panel (late snapshots, right). Thus, in the case of migration, we can estimate the secular evolution using a constant $a_N$. When $e_N$ or $i_N$ has not yet damped, we can estimate the secular evolution as if $a_N$ were constant at the location where Neptune spent most time. Because we have chosen a migration law for which $\frac{\dot{a}_N}{a_N}$ decreases with time, planetesimals behave as if $a_N$ has always had the value $a_N(t)$. We will discuss migration in detail in Section \ref{subsec:migrate}.

\begin{figure*}[t!]
\begin{center}
\includegraphics[width=\textwidth]{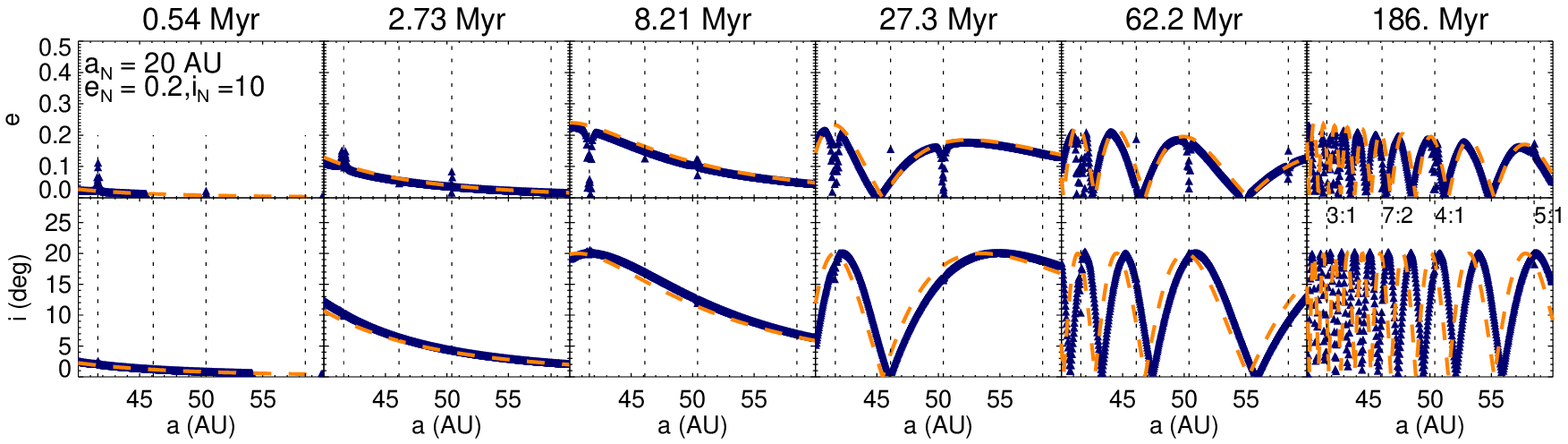}
\includegraphics[width=\textwidth]{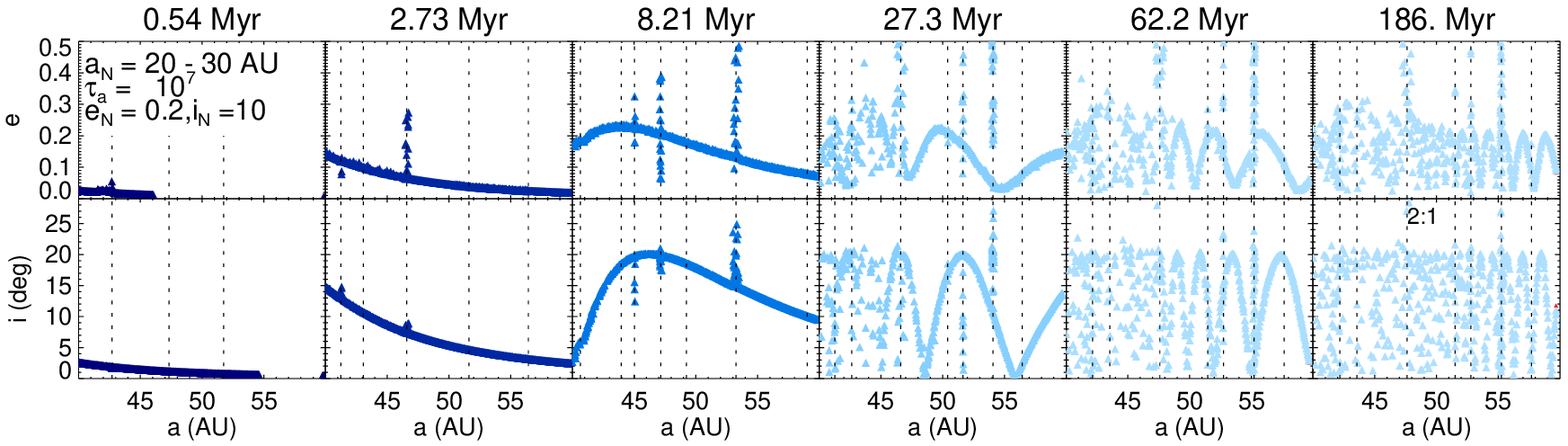}
\includegraphics[width=\textwidth]{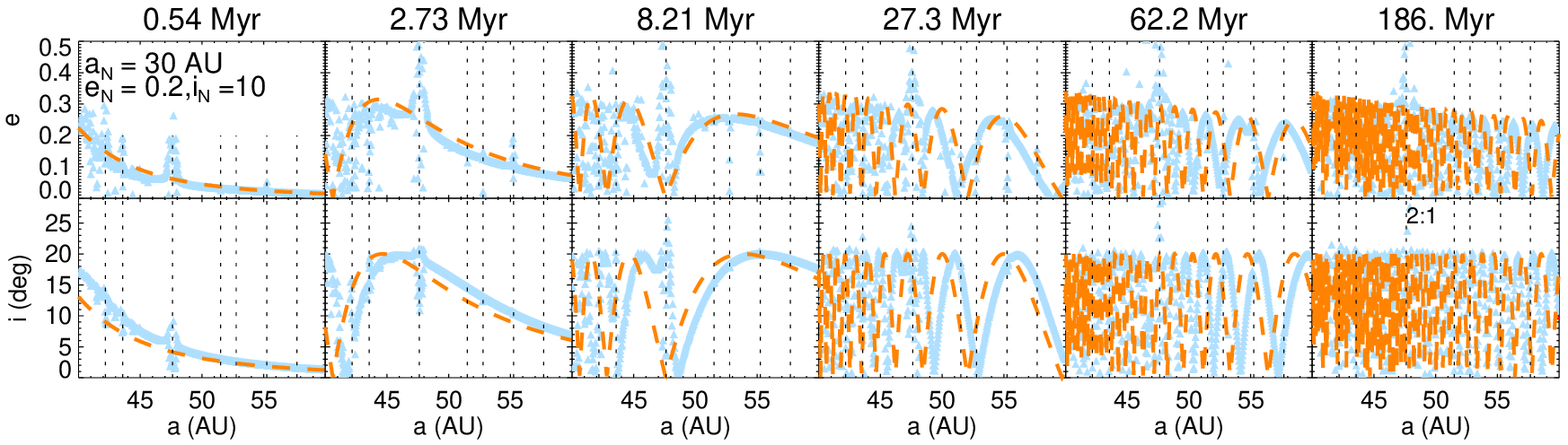}
\caption{Snapshots of the orbital evolution of test particles beginning with $e = i = 0$ from three different integrations (top, middle, bottom), each with $e_N = 0.2$ and $i_N = 10^\circ$ remaining constant. The top row of snapshots is from an integration in which Neptune's semi-major axis remains fixed at 20 AU; in the middle row, Neptune migrates from 20 to 30 AU on the timescale of $\tau_a = 10$ Myr; and in the final row Neptune's semi-major axis remains fixed at 30 AU. The dashed line is the predicted first-order secular evolution of the planetesimals (Eqn. \ref{eqn:sece} and \ref{eqn:seci}). The color of the planetesimals corresponds to the semi-major axis of Neptune, ranging from dark blue (20 AU) to light blue (30 AU). The planetesimals in the integration that includes migration of Neptune (middle panel) matches the top panel (Neptune at 20 AU) at the beginning (left snapshot) and the bottom panel (Neptune at 30 AU) at the end (right snapshot). \label{fig:excite}}
\end{center}
\end{figure*}

If Neptune's eccentricity and inclination are small (Fig. \ref{fig:obey}), the test particles remain below the observational limit of $e< 0.1, i< 6^\circ$ indefinitely. Since, to first order, $\iforced$ is independent of $\alpha$, the planetesimal's position relative to Neptune, the planetesimals will remain below $i < 6^\circ$ if $i_N < 3^\circ$. The forced eccentricity does depend on $\alpha$ (Fig. \ref{fig:freqs}), decreasing with the planetesimal's distance from Neptune. Thus to satisfy our conservative criteria established in Section \ref{sec:obs}, which requires planetesimals with $42.5 < a < 45$ AU to remain at $e < 0.1$, Neptune's eccentricity must stay below 0.09 at 20 AU or 0.06 at 30 AU. However, if $e_N$ and $i_N$ damp due to dynamical friction with the planetesimals, the initial values of Neptune's eccentricity and inclination can be higher. In the next subsection, we discuss the effects of damping.

\begin{figure}[htbp]
\begin{center}
\includegraphics[width=.45\textwidth]{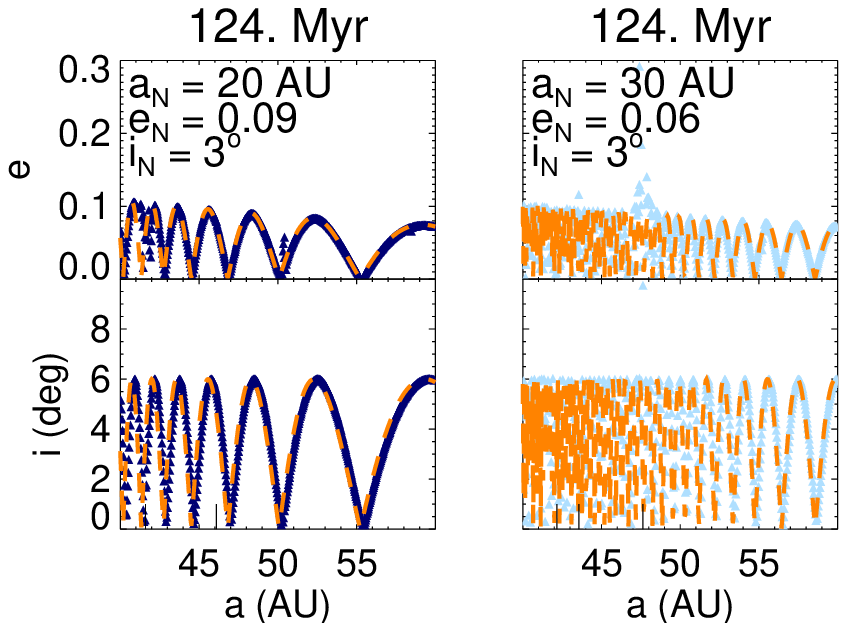}
\caption{A single snapshot, at 124 Myr, of test particles beginning with $e = i = 0$ from two integrations (left and right) with the Neptune's $a_N, e_N, i_N$ remaining constant. The integrations are identical except Neptune has $a_N = 20, e_N = 0.09$ in the left panel and $a_N = 30, e_N = 0.06$ in the right panel. Neptune has $i_N = 3^\circ$ in both panels. In both cases, the planetesimals are confined to the low eccentricities and inclinations established in Section \ref{sec:obs}. The dashed line is the predicted first-order secular evolution of the planetesimals (Eqn. \ref{eqn:sece} and Eqn. \ref{eqn:seci}). \label{fig:obey}}
\end{center}
\end{figure}

\subsection{Damping of Neptune's eccentricity and inclination}
\label{subsec:damp}

The constraints we placed previously were on how high Neptune's eccentricity and inclination can remain indefinitely. During secular evolution, planetesimals reach eccentricities up to $2 \eforced$ and inclinations up to $2\eforced$. However the planetesimals evolve to final values less than these maximum values, when Neptune's eccentricity and inclination damp. Here we refine the limits on Neptune's eccentricity and inclination in light of damping, considering a range of timescales for the dynamical-friction-driven damping rates of $e_N$ and $i_N$. The actual damping timescales depend on the surface density and size distribution of the planetesimals. 

\subsubsection{Eccentricity and inclination can be treated separately}
To first order, the effects on the planetesimals of Neptune's inclination and eccentricity, including damping, can be treated independently. In first-order secular theory (Section \ref{subsec:sec}), the evolution of a planetesimal's $e$ is independent of Neptune's inclination $i_N$ and of a planetesimal's $i$ is independent of Neptune's eccentricity $e_N$. Thus we can place constraints on the evolution of Neptune's eccentricity without taking into account its inclination or vice versa. Fig. \ref{fig:dampboth} illustrates the independence of the eccentricity and inclination parameters.

\begin{figure}[htbp]
\begin{center}
\includegraphics[width=.45\textwidth]{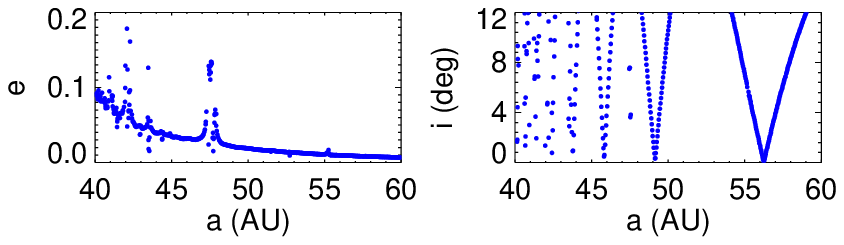}
\includegraphics[width=.45\textwidth]{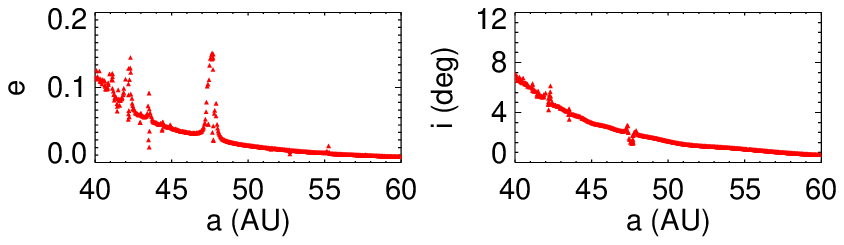}
\includegraphics[width=.45\textwidth]{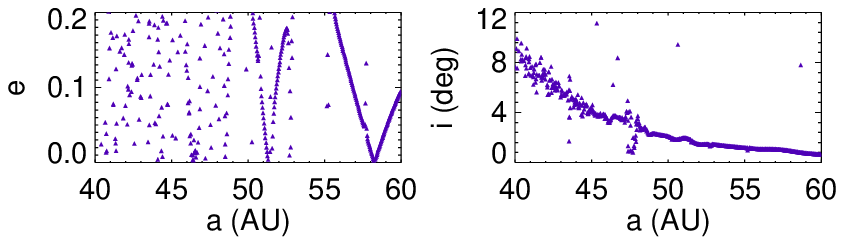}
\caption{The eccentricity and inclination damping of Neptune have a separable effect on the planetesimals. Snapshots at 22 Myr from three different integrations (top, middle, bottom) of test particles beginning with $e = i = 0$ with Neptune at $a_N = 30, e_N = 0.2, i_N = 10$. In the top panel (blue), Neptune's eccentricity damps on a timescale of 0.3 Myr and its inclination remains constant. In the bottom panel (purple), Neptune's inclination damps on a timescale of 0.3 Myr and its eccentricity remains constant. In the middle panel (red), both the eccentricity and inclination damp on a timescale of 0.3 Myr. In the middle planel, eccentricities of the planetesimals match the case in which just the eccentricity of Neptune damps (top) and the inclinations match the case in which just the inclination of Neptune damps (bottom). \label{fig:dampboth}}
\end{center}
\end{figure}

\subsubsection{Two regimes for damping: slow and fast}

The damping of Neptune's eccentricity $e_N$ or inclination $i_N$ affects the secular excitation of the planetesimals in two regimes: slow and fast. We summarize constraints on Neptune's eccentricity and inclination in Table \ref{tab:constrain}

\begin{table}[hbpt]
\caption{Constraints on Neptune's eccentricity and inclination \label{tab:constrain}}
\begin{tabular}{l|lll}
\\
			&no damping					&slow damping			&fast damping\\
			\hline
$e_N$		&$2 \alpharat e_{N} < 0.1$	&$\alpharat e_{N} < 0.1$	&$\alpharat e_{N} \sin(\gkbo \tau_e) < 0.1$	\\\\
$i_N$		&$2 i_N < 6^\circ$			&$ i_N < 6^\circ$		&$ i_N \sin(\gkbo \tau_i) < 6^\circ$	\\
\end{tabular}
\end{table}

In the slow regime, $e_N$ or $i_N$ damps on a timescale $\tau_e$ or $\tau_i$, respectively, longer than the time ($\frac{1}{2} 2\pi/g_{\rm KBO}$) for $e$ or $i$ to reach its maximum value. Because each planetesimal has an initial $i (t=0) = e (t=0)= 0$, its free inclination $\ifree$ and eccentricity $\efree$ are set by Neptune's initial $e_N (t=0)$ and $i_N (t=0)$. In this slow regime, the planetesimal's $\ifree$ and $\efree$ are conserved. As $e_N$ and $i_N$ damp, the planetesimal's forced eccentricity $\eforced$ and inclination $\iforced$ decrease, and the total eccentricity $e$ and inclination $i$ of the planetesimal approach $\efree$ and $\ifree$.

Thus, in the slow regime, the planetesimals will evolve to a value below $i < 6^\circ$ if $i_N < 6^\circ$. To satisfy our conservative criteria established in Section \ref{sec:obs}, which requires planetesimals with $42.5 < a < 45$ AU to remain at $e<0.1$, Neptune's eccentricity must stay below 0.18 at 20 AU and 0.12 at 30 AU. Thus in the slow damping case, there is a strong constraint (Table \ref{tab:constrain}) on the maximum eccentricity and inclination at which Neptune can remain over long timescales. Note that these values are twice the values given in Section \ref{subsec:evolve}. This is because in the slow damping regime, the planetesimals will evolve to $\efree = \eforced(t=0)$ and $\ifree = \iforced(t=0)$, whereas if $e_N$ and $i_N$ remain high indefinitely, the planetesimals' $e$ and $i$ continue to oscillate, reaching a maximum of $2\eforced$ and $2\iforced$.

If $e_N$ and $i_N$ damp in the fast regime, on a timescale shorter than the secular excitation time, $\ifree$ and $\efree$ are not conserved. The planetesimal's total $e$ and $i$ are frozen at the values they reach after approximately one damping time. Fig. \ref{fig:idamp} illustrates the behavior of the planetesimals in this regime in integrations in which Neptune's eccentricity or inclination damps. 

\begin{figure*}[hbtp]
\begin{center}
\includegraphics[width=.5\textwidth]{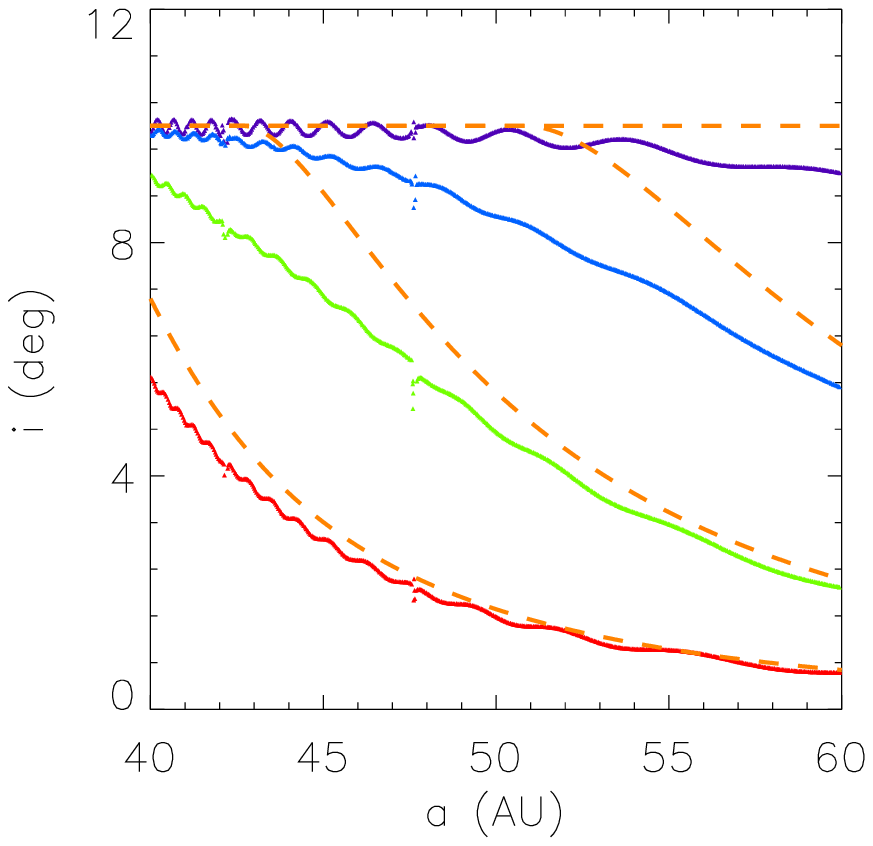}\includegraphics[width=.5\textwidth]{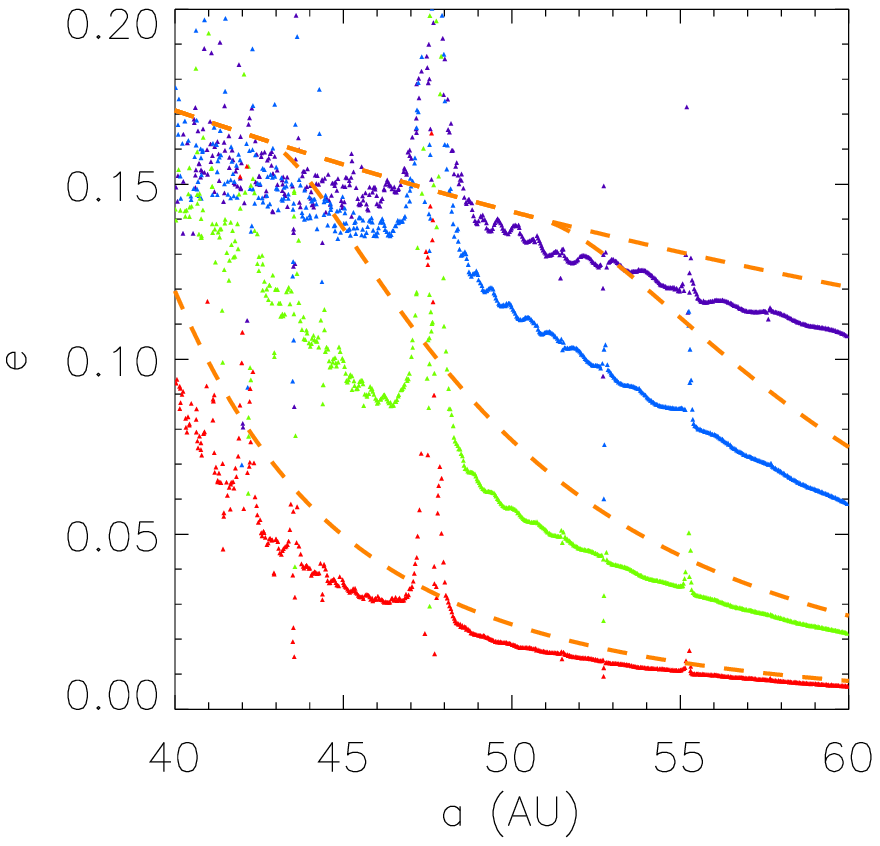}
\caption{Left: Snapshot at 44 Myr of the eccentricities and inclinations of test particles that began with $e = i = 0$. Neptune begins at $a_N = 30, e_N = 0, i_N = 10$. Then, in each of five different integrations, $i_N$ damps on a different timescale: 10 Myr (purple, top), 3 Myr (blue, second from top), 1 Myr (green, second from bottom), and 0.3 Myr (red, bottom). When the damping timescale is longer than the secular evolution timescale (purple; blue particles interior to 47 AU), the planetesimals damp to their initial free inclination, which is set by Neptune's initial inclination. When the damping timescale is shorter (red; green; blue particles beyond 47 AU), the planetesimals are frozen at the inclinations they reached at approximately the inclination damping timescale. Right: Same for eccentricity damping, with $a_N = 30, e_N = 0.2, i_N = 0$. The final eccentricity depends on the particle's location even in the slow damping regime (purple) because the initial forced eccentricity, the value to which the planetesimal's eccentricity converges, is a function of $\alpha$, the planetesimal's semi-major axis relative to Neptune (Eqn. \ref{eqn:simples}). In contrast, the forced inclination is independent of the particle's semi-major axis (Fig. \ref{fig:freqs}). Analytical curves are overplotted as orange dashed lines.\label{fig:idamp}}
\end{center}
\end{figure*}

Neptune's orbit could have been even more eccentric and inclined than in the slow regime if $e_N$ and $i_N$ damped quickly (Table \ref{tab:constrain}, right column). If the planetesimal has not yet reached the maximum of its secular cycle after one damping time, instead of converging to the $\efree$ and $\ifree$ set by initial conditions, the planetesimal evolves to $e_{\rm final} = \efree \sin(\gkbo \tau_e )$ and $i_{\rm final} =\ifree  \sin(\gkbo \tau_i )$ \citep[see][for a detailed explanation and justification]{2012D}. Consider again a planetesimal at 42.5 AU. If Neptune's eccentricity and inclination damp on a timescale of $\tau = $ 0.32 Myr, the final eccentricity and inclination of the planetesimal are reduced by a factor of $\sin(\gkbo \tau) = 0.5$ compared to the slow damping case. Thus the planetesimals will evolve to a value below $i < 6^\circ$ if $i_N < 6^\circ/0.5 = 12^\circ$. To satisfy our conservative criteria established in Section \ref{sec:obs}, which requires planetesimals with $42.5 < a < 45$ AU to remain at $e < 0.1$, $e_N$ must stay below $0.18/0.5 = 0.36$ at 20 AU and $0.12/0.5 = 0.24$ at 30 AU. 

We note that because ``slow" and ``fast" damping are defined relative to the secular evolution time, and because the secular evolution time increases with the planetesimal's semi-major axis, it may be that the damping is ``fast" for particles in the outer disk yet ``slow" for particles in the inner disk.

\subsection{Migration of Neptune}
\label{subsec:migrate}

\begin{figure}[htbp]
\begin{center}
\includegraphics[width=.45\textwidth]{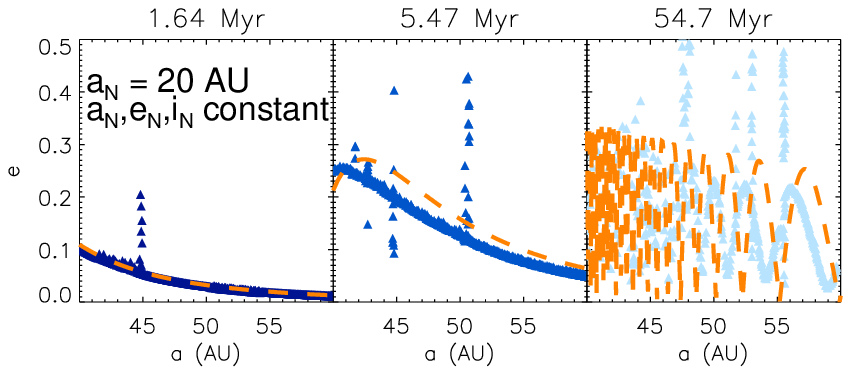}\\
\includegraphics[width=.45\textwidth]{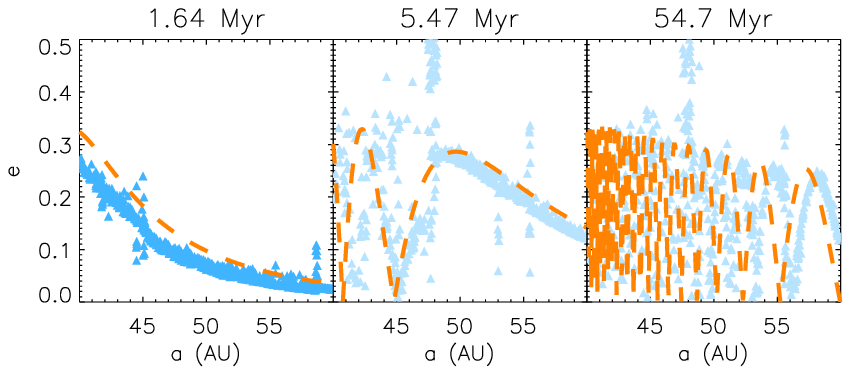}
\caption{Snapshots of the orbital evolution of test particles (blue) beginning with $e = i = 0$ from two different integrations (top, bottom), each with $e_N = 0.2, i_N = 1.77^\circ$ remaining constant and Neptune migrating from 20 AU to 30 AU. In the top model, the migration timescale is 10 Myr and in the bottom model the migration timescale is 1 Myr. The secular evolution model (dashed line) is computed at Neptune's location at the time of the snapshot, as if Neptune had been there during the entire secular evolution. \label{fig:mig2}}
\end{center}
\end{figure}

\begin{figure*}[htbp]
\begin{center}
\includegraphics[width=.45\textwidth]{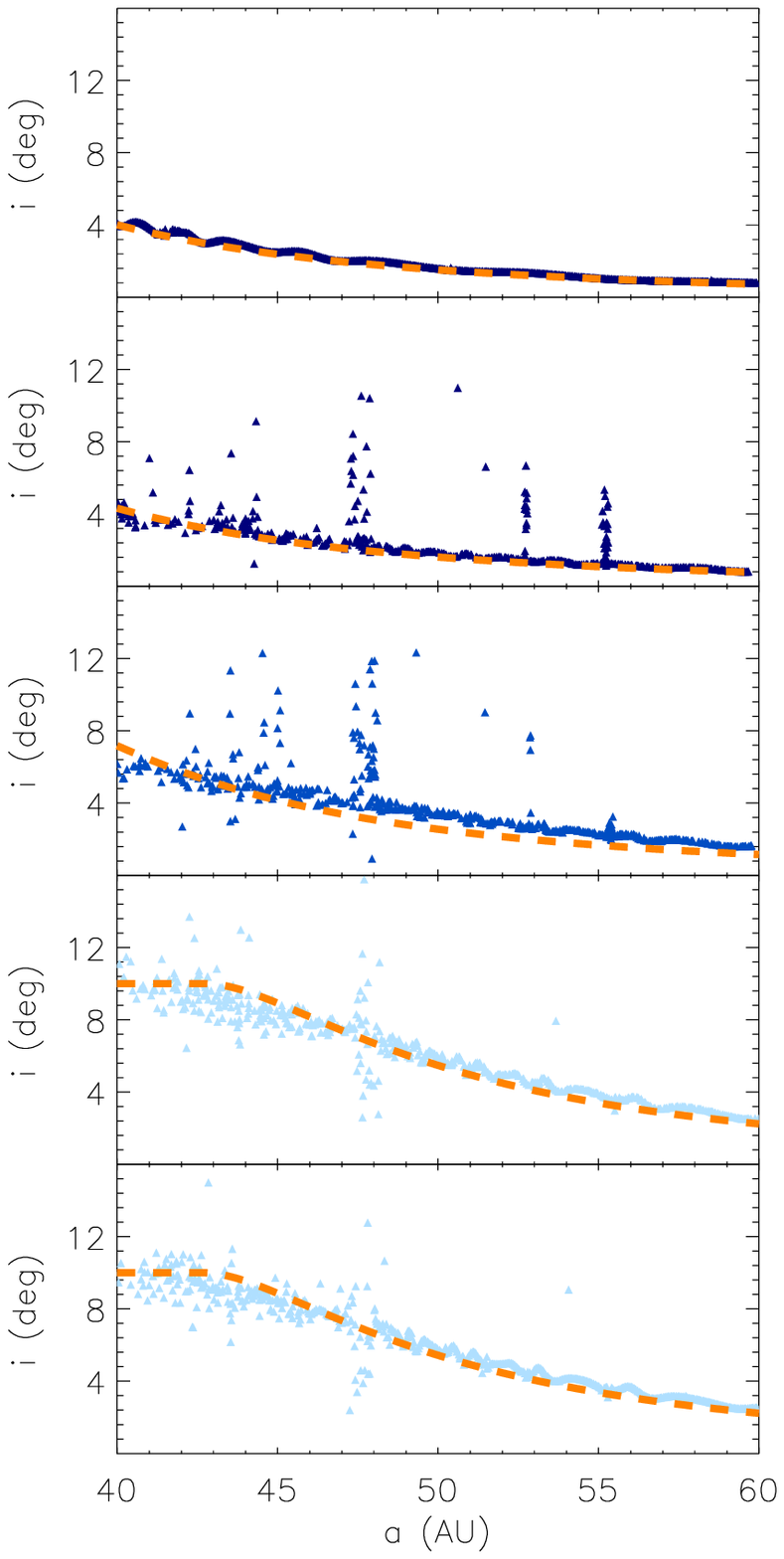} \includegraphics[width=.45\textwidth]{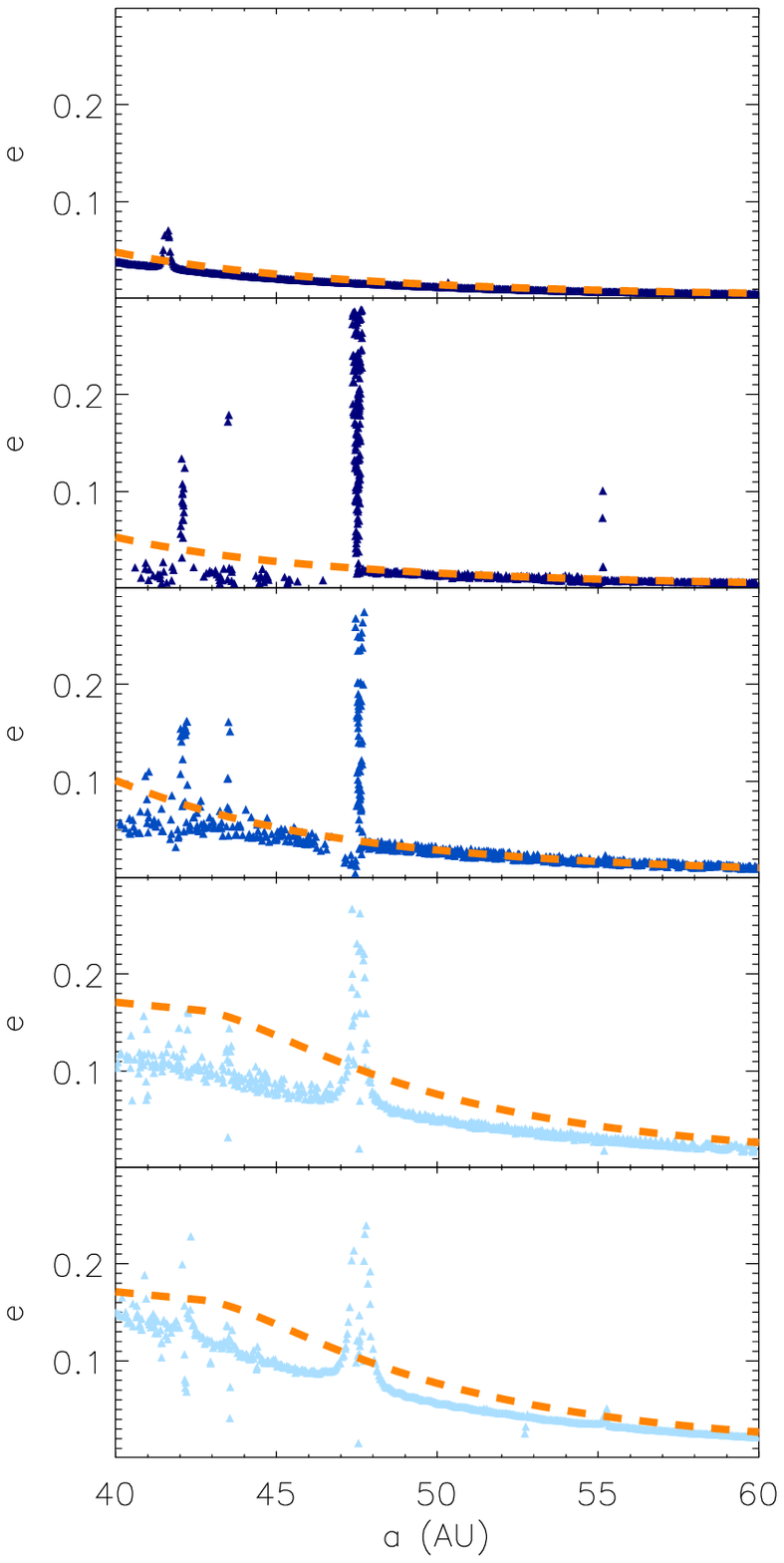}
\caption{Snapshots at 54.7 Myr from 10 different integrations. Left: Effect of the migration timescale when Neptune's inclination damps. In each case, Neptune has $e_N = 0$, an initial $i_N = 10$, and an inclination damping timescale of $\tau_i = 1$ Myr. In this fast-damping regime, Neptune's inclination damping timescale is shorter than the planetesimals' secular evolution timescale, so the inclinations of the planetesimals are frozen at approximately the values they reach at $\tau_i = 1$ Myr. In the top row, Neptune's semi-major axis stays fixed at 20 AU (infinitely slow migration). In rows 2-4, Neptune's migration timescale is 10 Myr, 3 Myr, and 1 Myr respectively. In row 5, Neptune's semi-major axis stays fixed at 30 AU (infinitely fast migration). The color of the planetesimals corresponds to Neptune's semi-major axis at the time of $\tau_i/2 = 0.5$ Myr, half the damping timescale, ranging from dark blue (20 AU) to light blue (30 AU). Because Neptune's semi-major axis sets the secular evolution time, the inclination that a given planetesimal reaches before it is frozen depends on the migration timescale relative to the inclination damping timescale. In rows 1-2, the model (dashed line) is computed with $a_N =  20$ AU. In the middle row, it is computed with $a_N = 24$ AU. In the rows 4-5, it is computed with $a_N = 30$ AU. Right: Same for eccentricity damping: in each case Neptune has $e_N = 0.2 $, an initial $i_N = 0$, and an eccentricity damping timescale of $\tau_e = 1$ Myr. \label{fig:dampmig}}
\end{center}
\end{figure*}

In the process of secular excitation of the cold classicals, migration alters a planetesimal's secular evolution timescale and forced eccentricity, which depend on $\alpha$, the ratio of the planetesimal's semi-major axis to Neptune's (the forced inclination is independent $\alpha$). For a planetesimal at 42.5 AU, the forced eccentricity is 40$\%$ larger when Neptune is at 30 AU vs. 20 AU and the secular evolution period 5 times shorter. Fig. \ref{fig:mig2} shows the effect of migration on the secular evolution of the planetesimals. The excitation at a given time is well-modeled by the secular theory (Eqn. \ref{eqn:sece} and \ref{eqn:seci}) using Neptune at its snapshot location. Recall that we are using a migration law that slows with time, so that Neptune spends increasing amounts of time at each, subsequent location.

For retaining the cold classicals, what matters is the ratio of the migration timescale to the damping timescale. When this ratio is large (``slow" migration), Neptune effectively damps at its initial position and we can model the secular evolution at this location (Fig. \ref{fig:dampmig}, top two rows). When this ratio is small (``fast" migration), Neptune effectively damps at its final position and we can model the secular evolution there (Fig. \ref{fig:dampmig}, bottom two rows). When the ratio is order unity, modeling the secular evolution at the location Neptune reaches after half a damping time is a decent approximation (Fig. \ref{fig:dampmig}, middle row). However, we have not explored the regime in which $\tau_a \sim \tau_e$ or $\tau_a \sim \tau_i$ in detail. Typically, we expect the damping and migration to occur in either the fast or slow regime, for reasons we will now state. The models in which Neptune is scattered from its location of formation to close to its current location \citep[e.g.][]{2008L} can be considered fast migration. For extensive migration in a planetesimal disk, we expect the migration timescale to be significantly longer than the damping timescale because the ``random momentum" is a fraction of order of $e$ or $i$ the Keplerian momentum. (Though we note that Neptune only has to migrate some fraction of its semi-major axis.)

\subsection{Two example scenarios}
\label{subsec:ex}

\begin{figure*}[htbp]
\begin{center}
\includegraphics[width=\textwidth]{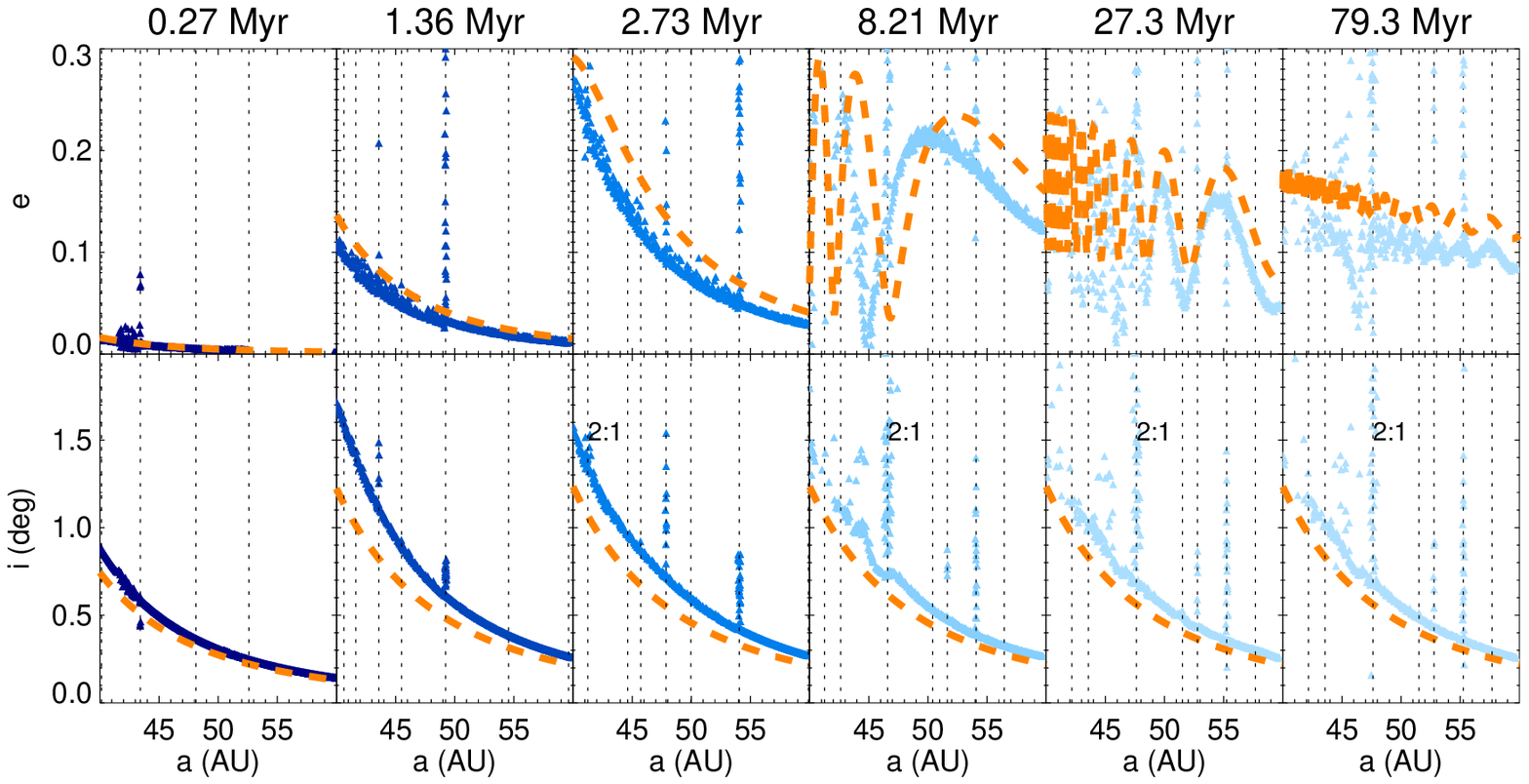}
\caption{In this example integration, the semi-major axis, inclination, and eccentricity of Neptune evolve on three different timescales: $\tau_a = 3$ Myr$, \tau_i = 0.3$ Myr$, \tau_e = 30$ Myr. The initial conditions of Neptune are $a_N =$ 20 AU, $e_N = 0.2, i_N = 10^\circ$. In each snapshot, the color of the planetesimals corresponds to Neptune's semi-major axis, from 20 AU (dark blue) to 30 AU (light blue). For the eccentricity, the model (dashed line) is calculated at the location of Neptune in the particular snapshot. For the inclination, because the inclination damping is fast compared to the migration, the model (dashed line) is calculated at Neptune's initial location of 20 AU. \label{fig:ex1}}
\end{center}
\end{figure*}

\begin{figure*}[htbp]
\begin{center}
\includegraphics[width=\textwidth]{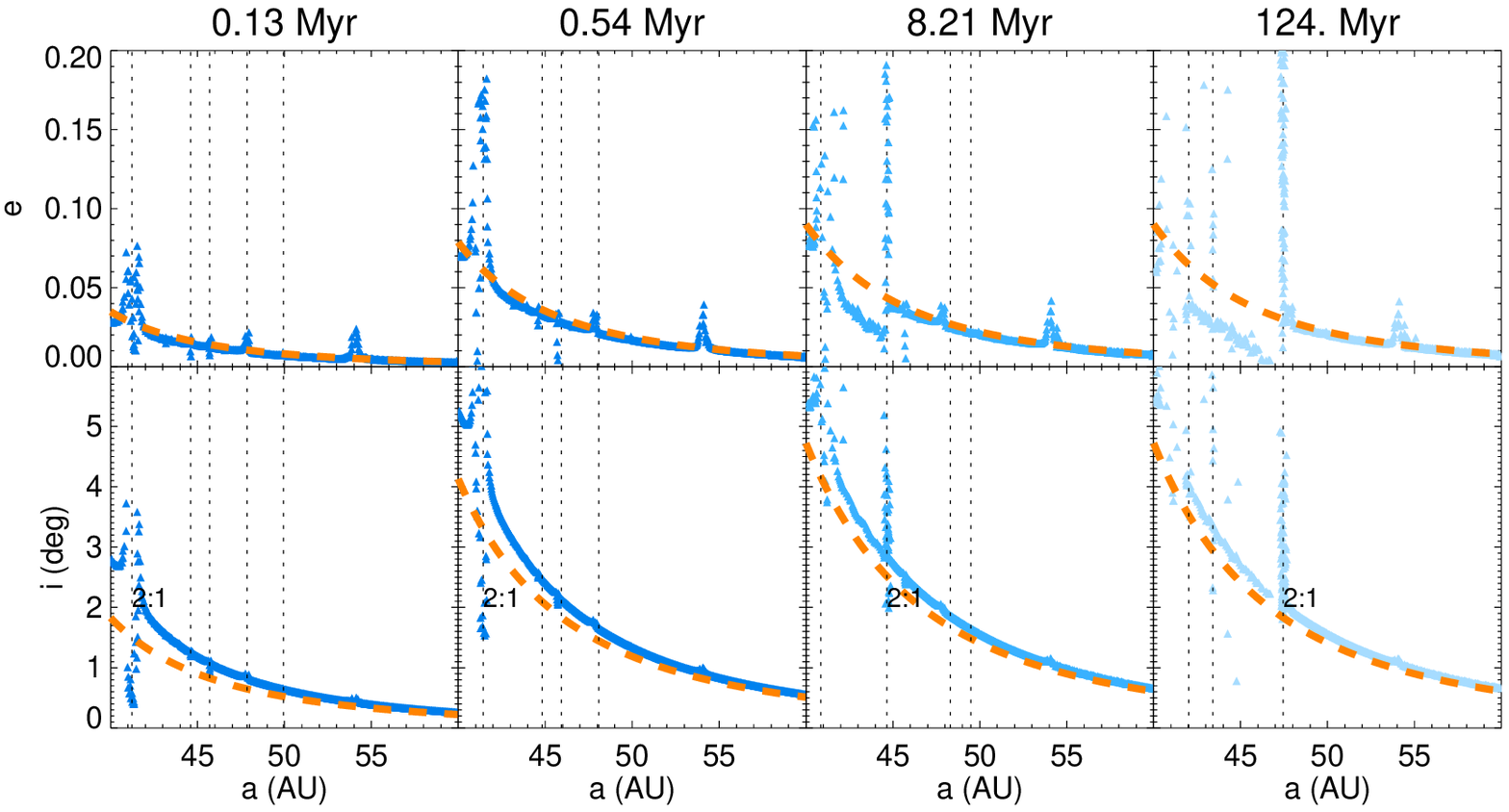}
\caption{Example integration for a region of the parameter space of Neptune's orbital evolution that is consistent with producing the observed cold classical population. Neptune migrates from 26 to 30 AU on a timescale of $\tau_a = 10$ Myr. Its initial eccentricity and inclination are 0.35 and 14$^\circ$ respectively and both damp on a timescale of $\tau_e = \tau_i = 0.3$ Myr. In each snapshot, the color of the planetesimals corresponds to Neptune's semi-major axis, from 26 AU (medium blue) to 30 AU (light blue). The model (dashed line) is calculated at the location of Neptune in the particular snapshot. \label{fig:ex2}}
\end{center}
\end{figure*}

Consider two example dynamical histories for Neptune. In each, the planet starts at an initial location, eccentricity, and inclination and undergoes migration, eccentricity damping, and inclination damping to evolve to its current orbit. First we consider a pathological scenario in which the evolution of the Neptune's semi-major axis, eccentricity, and inclination occur each on a different timescale (Fig. \ref{fig:ex1}). Neptune begins at $a_N$ = 20 AU, $e_N = 0.2$, and $i_N = 10 ^\circ$. On a timescale of $\tau_a = 3$ Myr, it migrates to its current location at 30 AU. The eccentricity damps on a longer timescale of $\tau_e = $30 Myr and the inclination on a shorter timescale of $\tau_i = 0.3$ Myr. For the inclination damping, the migration occurs in the slow regime $(\tau_a/\tau_i > 1)$, and we model the damping as taking place at 20 AU. With Neptune at 20 AU, the secular timescale of a planetesimal at 45 AU is $\frac{2\pi}{\gkbo} = $ 24 Myr and the initial forced inclination of the planetesimal is $\iforced = i_N  = 10^\circ$. The value of inclination to which the planetesimal evolves is reduced by a factor of $\sin(\gkbo \tau_i) = 0.07$. Thus the planetesimal at 45 AU evolves to a final value of $0.7^\circ$. For the eccentricity, the migration occurs in the fast regime. Because the eccentricity damping timescale is much longer than the migration timescale (fast migration), we model the eccentricity damping as taking place at 30 AU. With Neptune at 30 AU, the secular timescale of a planetesimal at 45 AU is $\frac{2\pi}{\gkbo} = $ 6 Myr and its initial forced eccentricity is $\eforced = 0.78 e_N = 0.16$. The eccentricity damping is in the slow-damping regime, so the planetesimal's final eccentricity evolves to its initial forced eccentricity, $e = 0.16$. This predicted value and those for planetesimals at other semi-major axes (orange curve, Fig. \ref{fig:ex1}) match the integrations well, validating our method.

Finally, we consider a more realistic scenario consistent with preserving the cold population (Fig. \ref{fig:ex2}). Neptune begins at 26 AU, $e_N = 0.35$, and $i_N = 14 ^\circ$. On a timescale of $\tau_a = 10$ Myr, it migrates to its current location at 30 AU. Its eccentricity and inclination damp on the timescale of $\tau_e = \tau_i = 0.3$ Myr. In this slow migration regime, we model the damping taking place with Neptune at 26 AU. Thus the secular timescale of a planetesimal at 45 AU is $\frac{2\pi}{\gkbo} \sim$11 Myr, its initial forced eccentricity is $0.7 e_N = 0.24$ and its initial forced inclination is $i_N  = 14^\circ$. The values of eccentricity and inclination to which the planetesimal evolves are reduced by a factor of $\sin(\gkbo \tau_e) = \sin(\gkbo \tau_i) = 0.16$. Thus the planetesimal at 45 AU evolves to a final eccentricity of $0.04$ and inclination of $2.4^\circ$, values satisfying the observational constraints established in Section \ref{sec:obs}.

\section{Conclusions}
\label{sec:conclusions}

As a first step in a comprehensive study of the impact on the Kuiper belt of a wide range of possible dynamical histories of the outer solar system, we have performed a suite of numerical integrations probing the impact of the orbital evolution of a single a planet on a disk of planetesimals.  We have presented the observational evidence for a population of dynamically cold objects in the Kuiper belt in the region from 42.5 to 45 AU that are confined to $e < 0.1$ and $i < 6^\circ$. We argued that recent models of Kuiper belt sculpting -- which explain many of the observed dynamical and physical properties of the Kuiper Belt -- do not generate or preserve sufficiently low eccentricities for the cold classicals \citep[e.g.][]{2003G,2008L,2008M}. Our results have revealed several principles key to constraining which orbital histories of Neptune are consistent with preserving the \emph{in situ} planetesimal population at the low eccentricities and inclinations required by the observations:

\begin{itemize}
\item If Neptune is scattered onto an eccentric and/or inclined orbit, it will secularly excite the eccentricities and inclinations of an \emph{in situ} planetesimal population.
\item Planetesimals starting with $e = i = 0$ reach eccentricities and inclinations up to twice their forced eccentricity and inclination, on timescales given by Eqn. (\ref{eqn:sece}) and (\ref{eqn:seci}).
\item As Neptune's eccentricity and inclination damp, the planetesimals evolve to their final eccentricities and inclinations. If the damping timescales $\tau_e$ and $\tau_i$ of Neptune's orbit are slow compared to the secular evolution time, a planetesimal evolves to its initial free eccentricity and inclination, which are set by Neptune's initial eccentricity and inclination. If the damping is fast compared to the secular evolution time, the planetesimal's $e$ and $i$ effectively freeze at the values they reach after one damping time, reaching final values of $\efree \sin(\gkbo \tau_e)$ and $\eforced \sin(\gkbo \tau_i)$ respectively. See Table \ref{tab:constrain} for constraints on Neptune's eccentricity and inclination in the two damping regimes.
\item The effects of Neptune's: 1) eccentricity evolution, and 2) inclination evolution, on the planetesimals can be treated separately to first order.
\item At a given location in the planetesimal disk, the secular excitation timescales and forced eccentricity (but not the forced inclination) depend on Neptune's location, which is altered by Neptune's migration. When Neptune's migration is slow relative to the damping time, the secular evolution effectively takes place at Neptune's initial location. When Neptune's migration is fast relative to the damping time, the secular evolution effectively takes place at Neptune's final location.
\end{itemize}

From these principles, it is evident that the three models described in \citet{2008L} would not be able to retain a cold classical population. Neptune begins with an eccentricity of 0.3 and a semi-major axis of 27.5 AU (run A and C) or 28.9 AU (run B) and damps on a timescale of 1 Myr (run A and B) or 3 Myr (run C), in the slow migration regime. When Neptune is at 27.5 AU, a planetesimal at 42.5 AU has a forced eccentricity of $e = 0.76 e_N = 0.23$ and a secular evolution timescale of 6 Myr. The final eccentricity of the planetesimal evolves to is reduced by a factor of $\sin(\gkbo \tau_e) = 0.85$ for a damping timescale of 1 Myr (fast damping) and is not reduced for a damping timescale of 3 Myr (slow damping). Thus the final eccentricity of the planetesimal is 0.19 (run A) or 0.23 (run C), well above the observational limit. When Neptune is at 28.9 AU (run B), a planetesimal at 42.5 AU has a forced eccentricity of $e = 0.79 e_N = 0.24$ and a secular evolution timescale of 5 Myr. The final eccentricity the planetesimal evolves to is reduced by a factor of $\sin(\gkbo \tau_e) = 0.97$ for a damping timescale of 1 Myr. Thus the final eccentricity of the planetesimal is 0.23, well above the observational limit. According to the constraints established in our paper, any of these three initial conditions for Neptune could retain the cold classicals if Neptune's eccentricity were to damp more quickly. 

Based on these principles, we can place robust constraints on Neptune's dynamical history. For example, in the regime of slow damping and slow migration, if Neptune's initial semi-major axis is 20 AU, then its eccentricity must stay below 0.18 and inclination below 6 degrees. If Neptune's initial semi-major axis is 30 AU -- or if it migrates quickly, relative to the damping time, from its initial location to 30 AU -- then its eccentricity must stay below 0.12. In the case of fast damping -- on a timescale shorter than a planetesimal's secular excitation time -- the initial eccentricity and inclination of Neptune can be even higher.

Having established these principles, we complete our parameter study in two companion papers \citet{2012D} and Dawson and Murray-Clay (2012b), in prep. In the first companion paper, we consider more generally the constraints on Neptune's dynamical history from the eccentricity distribution of the classical KBOs and incorporate several other important effects in the constraints from the cold classicals:
\begin{itemize}
\item A more accurate model for secular excitation that includes higher-order terms.
\item The greatly increased secular frequency near Neptune's mean-motion resonances, which places strong constraints on Neptune's semi-major axis when its eccentricity is high.
\item The effects of the other giant planets, including precession of Neptune, which can reduce a planetesimal's forced eccentricity \citep{2011B}, and oscillations in Neptune's semi-major axis, which can create a chaotic sea in the Kuiper belt region.
\end{itemize}
Then we combine these constraints for retaining the cold classicals with constraints for creating the hot classical population and identify which regions of parameter space of Neptune's dynamical history can produce the hot classical population without disrupting the cold population. In the second companion paper, we place constraints on Neptune's inclination and inclination damping time.

Our parameterization of Neptune's orbital history allows us to constrain which orbital histories are consistent with maintaining the cold classical population. By combining the observational constraints (Sec. \ref{sec:obs}) with the principles derived in this paper, we can immediately check whether a particular set of initial conditions for Neptune's semi-major axis, eccentricity, and inclination, and the rates of its migration and eccentricity and inclination damping, are consistent with maintaining the observed dynamically cold population, without performing computationally expensive integrations. A picture, both qualitative and quantitative, is emerging of what dynamical histories of Neptune allow a promising general scenario -- Neptune's delivery of the hot classicals from the inner disk to classical region, where the cold population has formed \emph{in situ} -- to be consistent with observed low eccentricities and inclinations of the cold classical population. Barring fast precession of Neptune's orbit, which we do not consider in this work, the existence of the cold classical population implies that Neptune could have spent only a limited time at high eccentricity and/or inclination during its dynamical history.

\acknowledgments We thank an anonymous referee for helpful comments, including recommending the addition of Section 3.4. SW gratefully acknowledges funding by the National Science Foundation Research Experiences for Undergraduates (REU) and the Department of Defense Awards to Stimulate and Support Undergraduate Research Experiences (ASSURE) programs under Grant no. 0754568. RID is supported by a National Science Foundation Graduate Research Fellowship under grant DGE-1144152 and RMC by the Smithsonian Institution. The computations in this paper were run on the Odyssey cluster supported by the Harvard FAS Sciences Division Research Computing Group.

\appendix
\section{Modifications to the equations of motion for the evolution of Neptune's orbital elements}
\label{sec:migdamp}

In this appendix, we follow \citet{2002L} to derive modifications to the equations of motion of an orbiting body to allow any form of evolution of the body's orbital elements. We use these modified equations to model the early solar system evolution of Neptune's orbit caused by interactions with the other giant planets and with the planetesimal disk. Directly modeling Neptune, the other giant planets, and tens of thousands of massive planetesimals would be computationally prohibitive. Instead, we model Neptune alone, with its orbit evolving as it would under the influence of the other planets and the planetesimal disk, including migration, damping of its eccentricity, and damping of its inclination. However, the equations we derive can be used to implement any type of orbital evolution, not just migration and damping. For example, in a companion paper \citep{2012D}, we use the modifications to cause Neptune's semi-major axis to oscillate, as it would due to resonant interactions with Uranus.

We modify the \emph{Mercury 6.2} \citep{1999C} N-body integration code to allow for arbitrary orbital evolution by adding extra terms to the equations of motion at each timestep. We add these terms via a user-defined force, already supported by \emph{Mercury 6.2}, and an additional velocity modification described below. We note that \citet{2002L} use their derived modifications for a non-symplectic code and use a symplectic implementation when employing a symplectic algorithm. However, we do not add the extra terms in a symplectic manner, even for a symplectic integration algorithm; thus these modifications must represent small perturbations on the planet. We determined that a timestep of 200 days was sufficiently small for our models, yielding the same results as smaller step sizes. 

We make these modifications as follows:
\begin{enumerate}
\item At each timestep, \emph{Mercury} 6.2 calls the user-defined force routine. We have modified this routine to return not only a user-defined acceleration but a user-defined velocity. We also modify the routine to only apply these corrections for Neptune, not for the planetesimals.
\item The user-defined force routine calculates the user-defined velocity additional terms from Eqn. (\ref{xvel})-(\ref{zvel}) below and the user-defined acceleration additional terms from Eqn. (\ref{xacc})-(\ref{zacc}) below.
\item The acceleration additional terms are used to advance the velocity of Neptune and the velocity additional terms to advance the position of Neptune, in addition to the usual gravitational forces.
\end{enumerate}

To derive these additional terms, we follow \citet{2002L} but instead of holding the orbital inclination $i$ constant, we allow it to vary with time

Equation (\ref{position}) gives the position vector components ($x$, $y$, and $z$ ) as a function of the orbital elements (compare to Eqn. A5 of \citeauthor{2002L}). The variable $r$ is the distance of the planet from central body, $\Omega$ is the longitude of the ascending node in the xy plane, $\omega$ is the argument of periapse, $f$ is the true anomaly, and $i$ is the inclination.

\begin{align}
\label{position}
x &= r \cos{\Omega} \cos{(\omega + f)} - r \cos{i} \sin{\Omega} \sin{(\omega + f)} \notag \\
y &= r \sin{\Omega} \cos{(\omega + f)} + r \cos{i} \cos{\Omega} \sin{(\omega +f)}  \\ 
z &= r \sin{i} \sin{(\omega + f)} \notag \\ \notag
\end{align}

\noindent The velocity vector components are given in equation \eqref{velocity} (compare to Eqn. A6 of \citeauthor{2002L}).  A dot over a variable indicates its derivative with respect to time.
\begin{align}
\dot{x} &=  \cos{\Omega} \,[\dot{r} \cos{(\omega + f)} - r \dot{f} \sin{(\omega + f)}] - \sin{\Omega} \,[\dot{r} \cos{i} \sin{(\omega + f)} + r \dot{f} \cos{i} \cos{(\omega + f)} -z \dot{i}]  \notag \\ 
\label{velocity}
\dot{y} &=  \sin{\Omega} \,[\dot{r} \cos{(\omega + f)} - r \dot{f} \sin{(\omega + f)}] + \cos{\Omega} \,[\dot{r} \cos{i} \sin{(\omega + f)} + r \dot{f} \cos{i} \cos{(\omega + f)} - z \dot{i}]   \\ 
\dot{z} &=  \dot{r} \sin{i} \sin{(\omega + f)} + r \dot{f} \sin{i} \cos{(\omega + f)}]  +r \dot{i} \cos{i} \sin{(\omega + f)} \notag \\ \notag 
\end{align}

Below are the x-components of the ``velocity" and ``acceleration" additional terms used to update the body's position and velocity, respectively, at each timestep (compare to Eqn. A3 and A4 of \citeauthor{2002L}). The variable $a$ is the semi-major axis and $e$ is the orbital eccentricity.

\begin{align}
\label{xchange}
\frac{dx}{dt}  \bigg|_{\dot{a}} + \frac{dx}{dt}  \bigg|_{\dot{e}} + \frac{dx}{dt}  \bigg|_{\dot{i}} &= \frac{\partial x}{\partial a} \dot{a} +  \frac{\partial x}{\partial e} \dot{e} +  \frac{\partial x}{\partial i} \dot{i} \\ \notag \\
\label{vxchange}
\frac{d \dot{x}}{dt}  \bigg|_{\dot{a}} + \frac{d \dot{x}}{dt}  \bigg|_{\dot{e}} + \frac{d \dot{x}}{dt}  \bigg|_{\dot{i}} &= \frac{\partial \dot{x}}{\partial a} \dot{a} +  \frac{\partial \dot{x}}{\partial e} \dot{e} +  \frac{\partial \dot{x}}{\partial i} \dot{i} \\ \notag
\end{align}

Similar expressions can be derived for the other coordinates. It should be noted that these are simply the additional terms to the equations of motion resulting from the orbital evolution -- $\dot{a}$, $\dot{e}$, and/or $\dot{i}$ -- and do not describe the overall motion of the system. All coordinates, velocities, and accelerations must be calculated with respect to the central body. In order to compute these partial derivatives, we must first define the variables $r$, $\dot{r}$ and $r \dot{f}$ following \citet{2000M}:
 
\begin{equation}
\label{partials}
r = \frac{a (1 - e^{2})}{1 + e \cos{f}} \;  \; \; \;  \; \; \; \; \; \;  \dot{r} = \frac{n a}{\sqrt{1 - e^{2}}} e \sin{f} \;  \; \; \; \; \; \; \; \; \;  r \dot{f} =  \frac{n a}{\sqrt{1 - e^{2}}} (1 + e \cos{f}) \\
\end{equation}

Next we calculate the partial derives of $r, \dot{r},$ and $r\dot{f}$ with respect to the orbital elements $a$, $e$, and $i$ (compare to Eqn. A7 of \citeauthor{2002L}):

\begin{eqnarray}
\label{dr}
\frac{\partial r}{\partial a} = \frac{r}{a} \; \; \; \; \; \; \; \; \; \; \; \; \frac{\partial r}{\partial e} = \bigg[ - \frac{2 e r}{1 - e^{2}} - \frac{r^{2} \cos{f}}{a (1 - e^{2})}\bigg] \; \; \; \; \; \; \; \; \; \; \; \; \frac{\partial r}{\partial i} = 0 \\ \nonumber \\ 
\label{drdot}
\frac{\partial \dot{r}}{\partial a} = - \frac{\dot{r}}{2 a} \; \; \; \; \; \; \; \; \; \; \; \; \; \; \; \; \; \; \; \; \;   \frac{\partial \dot{r}}{\partial e} = \frac{\dot{r}}{e (1 - e^{2})}  \; \; \; \; \; \; \; \; \; \; \; \; \; \; \; \; \; \; \; \; \;   \frac{\partial \dot{r}}{\partial i} = 0 \\ \nonumber \\
\label{drfdot}
\frac{\partial (r \dot{f})}{\partial a} = - \frac{r \dot{f}}{2 a} \; \; \; \; \; \; \; \; \;  \frac{\partial (r \dot{f})}{\partial e} = \frac{r \dot{f} (e + \cos{f})}{(1 - e^{2}) (1 + e \cos{f})} \; \; \; \; \; \; \; \; \;  \frac{\partial (r \dot{f})}{\partial i} = 0  \\ \notag
\end{eqnarray}

Combining equations \eqref{dr} through \eqref{drfdot} with equation \eqref{xchange} yields an expression for change in position (compare to Eqn. A8 of \citeauthor{2002L}):

\begin{align}
\label{xvel}
\frac{d x}{d t} \bigg|_{\dot{a}} + \frac{d x}{d t} \bigg|_{\dot{e}} + \frac{d x}{d t} \bigg|_{\dot{i}} &= \frac{x}{a} \dot{a}+ \bigg[ \frac{r}{a (1 - e^{2})} - \frac{1 + e^{2}}{1 - e^{2}} \bigg] \frac{x}{e} \dot{e} + (z \, \sin{\Omega}) \, \dot{i} \\ \notag \\
\label{yvel}
\frac{d y}{d t} \bigg|_{\dot{a}} + \frac{d y}{d t} \bigg|_{\dot{e}} + \frac{d y}{d t} \bigg|_{\dot{i}} &= \frac{y}{a} \dot{a} + \bigg[ \frac{r}{a (1 - e^{2})} - \frac{1 + e^{2}}{1 - e^{2}} \bigg] \frac{y}{e} \dot{e} - (z \, \cos{\Omega}) \, \dot{i} \\ \notag \\
\label{zvel}
\frac{d z}{d t} \bigg|_{\dot{a}} + \frac{d z}{d t} \bigg|_{\dot{e}} + \frac{d z}{d t} \bigg|_{\dot{i}} &= \frac{z}{a} \dot{a} + \bigg[ \frac{r}{a (1 - e^{2})} - \frac{1 + e^{2}}{1 - e^{2}} \bigg] \frac{z}{e} \dot{e} + ( - \sin{\Omega} x + \cos{\Omega} y) \, \dot{i} \\ \notag
\end{align}

\noindent We can now use these additional terms to update the position of a body undergoing any arbitrary orbital evolution in $a$, $e$, and/or $i$, including migration, eccentricity damping, and inclination damping.  Below we combine \eqref{dr} through \eqref{drfdot} with equation \eqref{vxchange} to obtain the acceleration terms used to update the body's velocity (compare to Eqn. A9 of \citeauthor{2002L}):

\begin{align}
\label{xacc}
\frac{d \dot{x}}{dt}  \bigg|_{\dot{a}} + \frac{d \dot{x}}{dt}  \bigg|_{\dot{e}} + \frac{d \dot{x}}{dt}  \bigg|_{\dot{i}} &=  \frac{- \dot{x} + 3 z \dot{i} \sin{\Omega}}{2 a} \dot{a} + \bigg(\cos{\Omega} \, \bigg[ \frac{\partial \dot{r}}{\partial e} \cos{(\omega + f)} - \frac{\partial (r \dot{f})}{\partial e} \sin{(\omega + f)} \bigg]  \notag \\ &- \sin{\Omega} \bigg[ \frac{\partial \dot{r}}{\partial e} \cos{i} \sin{(\omega + f)} + \frac{\partial (r \dot{f})}{\partial e} \cos{i} \cos{(\omega + f)} - \dot{i} \frac{\partial r}{\partial e} \frac{z}{r} \bigg] \bigg) \dot{e} \notag \\ &+ \sin{\Omega} \bigg[ \dot{z} + \frac{\dot{i}}{i} z \bigg] \dot{i} \\
\label{yacc}
\frac{d \dot{y}}{dt}  \bigg|_{\dot{a}} + \frac{d \dot{y}}{dt}  \bigg|_{\dot{e}} + \frac{d \dot{y}}{dt}  \bigg|_{\dot{i}} &=  \frac{- \dot{y} - 3 z \dot{i} \cos{\Omega}}{2 a} \dot{a} + \bigg(\sin{\Omega} \, \bigg[ \frac{\partial \dot{r}}{\partial e} \cos{(\omega + f)} - \frac{\partial (r \dot{f})}{\partial e} \sin{(\omega + f)} \bigg]  \notag \\ &+ \cos{\Omega} \bigg[ \frac{\partial \dot{r}}{\partial e} \cos{i} \sin{(\omega + f)} + \frac{\partial (r \dot{f})}{\partial e} \cos{i}\cos{(\omega + f)} - \dot{i} \frac{\partial r}{\partial e}\frac{z}{r} \bigg] \bigg) \dot{e} \notag \\ &- \cos{\Omega} \bigg[ \dot{z} + \frac{\dot{i}}{i} z \bigg] \dot{i} \\
\label{zacc}
\frac{d \dot{z}}{dt}  \bigg|_{\dot{a}} + \frac{d \dot{z}}{dt}  \bigg|_{\dot{e}} + \frac{d \dot{z}}{dt}  \bigg|_{\dot{i}} &=  \bigg[ \frac{- \dot{r}}{2 a} \sin{i} \sin{(\omega + f)} + \frac{ - r \dot{f}}{2 a} \sin{i} \cos{(\omega + f)} + \frac{r}{a} \dot{i} \cos{i} \sin{(\omega + f)} \bigg] \dot{a} \notag \\
&+ \bigg[ \frac{\partial \dot{r}}{\partial e} \sin{i} \sin{(\omega + f)} + \frac{\partial (r \dot{f})}{\partial e} \sin{i} \cos{(\omega + f)} + \frac{\partial r }{\partial e}\dot{i} \cos{i} \sin{(\omega + f)} \bigg] \dot{e} \notag \\ &+ \bigg[ \dot{r} \cos{i} \sin{(\omega + f)} + r \dot{f} \cos{i} \cos{(\omega + f)} + r \frac{\dot{i}}{i} \cos{i} \sin{(\omega + f)} - \dot{i} z \bigg] \dot{i}
\end{align}

\noindent Arbitrary functions for the evolution of the orbital elements can be assigned to the $\dot{a}/a$, $\dot{e}/e$ and $\dot{i}/i$ terms. With the additional terms derived above (Eqn. \ref{xvel} - \ref{zacc}), the full motion of a body undergoing arbitrary evolution in semi-major axis, eccentricity, and/or inclination can be implemented. When the eccentricity $e$ reaches a small value, machine round-off error can cause the eccentricity damping terms to damp the eccentricity below 0, leading to failure of the integrator. Therefore did not apply the eccentricity damping terms if $e < 10^{-4}$.If the orbit of the body precesses, $\omega$ is no longer constant (apsidal precession) and/or $\Omega$ is no longer constant (nodal precession). Since in this paper we do not consider precession \citep[see][]{2011B,2012D}, we leave the additional modifications to the equations of motion to account for precession for future work.

\bibliography{ms} \bibliographystyle{apj}

\end{document}